\documentclass[aps,pre,reprint,superscriptaddress]{revtex4-1}

\usepackage{color}
\usepackage{bm}
\usepackage{latexsym}
\usepackage{amssymb}
\usepackage{amsmath}  
\usepackage{graphicx}
\usepackage{empheq} 

\usepackage{soul}  
\usepackage{empheq} 
\usepackage{tabulary}
\usepackage{tabularx}

\begin{document}


\title{Ultrafiltration of charge-stabilized dispersions at low salinity}


\author{Rafael Roa}
\email[]{rafael.roa@helmholtz-berlin.de}
\affiliation{Forschungszentrum J\"ulich, Institute of Complex Systems (ICS-3), 52425 J\"ulich, Germany}
\affiliation{Helmholtz-Zentrum Berlin, Soft Matter and Functional Materials, 14109 Berlin, Germany}

\author{Daniel Menne}
\affiliation{RWTH Aachen University, Chemical Process Engineering, 52064 Aachen, Germany}

\author{Jonas Riest}
\affiliation{Forschungszentrum J\"ulich, Institute of Complex Systems (ICS-3), 52425 J\"ulich, Germany}
\affiliation{J\"ulich-Aachen Research Alliance, JARA-Soft Matter}

\author{Pompilia Buzatu}
\affiliation{RWTH Aachen University, Chemical Process Engineering, 52064 Aachen, Germany}
\affiliation{DWI Leibniz Institute for Interactive Materials, 52074 Aachen, Germany}

\author{Emiliy K. Zholkovskiy}
\affiliation{Ukrainian Academy of Sciences, Institute of Bio-Colloid Chemistry, 03142 Kiev, Ukraine}

\author{Jan K. G. Dhont}
\affiliation{Forschungszentrum J\"ulich, Institute of Complex Systems (ICS-3), 52425 J\"ulich, Germany}
\affiliation{Physics Department, Heinrich-Heine Universit\"at D\"usseldorf, 40225 D\"usseldorf, Germany}
\affiliation{J\"ulich-Aachen Research Alliance, JARA-Soft Matter}

\author{Matthias Wessling}
\affiliation{RWTH Aachen University, Chemical Process Engineering, 52064 Aachen, Germany}
\affiliation{DWI Leibniz Institute for Interactive Materials, 52074 Aachen, Germany}
\affiliation{J\"ulich-Aachen Research Alliance, JARA-Soft Matter}

\author{Gerhard N\"agele}
\email[]{g.naegele@fz-juelich.de}
\affiliation{Forschungszentrum J\"ulich, Institute of Complex Systems (ICS-3), 52425 J\"ulich, Germany}
\affiliation{Physics Department, Heinrich-Heine Universit\"at D\"usseldorf, 40225 D\"usseldorf, Germany}
\affiliation{J\"ulich-Aachen Research Alliance, JARA-Soft Matter}

\date{\today}

\begin{abstract}

We present a comprehensive study of cross-flow ultrafiltration (UF) of charge-stabilized suspensions, under low-salinity conditions of electrostatically strongly repelling colloidal particles. The axially varying permeate flux, near-membrane concentration-polarization (CP) layer and osmotic pressure profiles are calculated using a macroscopic diffusion-advection boundary layer method, and are compared with filtration experiments on aqueous suspensions of charge-stabilized silica particles. The theoretical description based on the one-component macroion fluid  
model (OCM) accounts for the strong influence of surface-released counterions on the renormalized colloid charge and suspension osmotic compressibility, and for the influence of the colloidal hydrodynamic interactions and electric double layer repulsion on the concentration-dependent suspension viscosity $\eta$, and collective diffusion coefficient $D_c$. A strong electro-hydrodynamic enhancement of $D_c$ and $\eta$, and likewise of the osmotic pressure, is predicted theoretically, as compared with their values for a hard-sphere suspension. We also point to the failure of generalized Stokes-Einstein relations describing reciprocal relations between $D_c$ and $\eta$. According to our filtration model,  $D_c$ is of dominant influence, giving rise to an only weakly developed CP layer having practically no effect on the  permeate flux.   
This prediction is quantitatively confirmed by our UF measurements of the permeate flux using an aqueous suspension of charged silica spheres as the feed system.  
The experimentally detected fouling for the largest considered transmembrane pressure values is shown  
not to be due to filter cake formation by crystallization or vitrification.
\end{abstract}


\maketitle

\section{Introduction}

The concentration and purification of dispersions of charged particles is required in many industrial, pharmaceutical and biological applications. For the concentration of particles strongly affected by Brownian motion, ultrafiltration (UF) has become a standard method whose major advantage is its low energy requirement. It is extensively used for the purification and concentration of a large variety of proteins \cite{Shen:1977hm,Bowen:1995ft,Bhattacharjee:1999cz,Bowen:2001ib,Kim:2006en,Bowen:2007ix,Rohani:2010bm,De:2011ef,Sarkar:2010du}, food stuff \cite{Jonsson:1984ee}, industrial enzymes and antibody fragments \cite{Dayal:2005,Lebreton:2008}, and waste water treatment \cite{Shannon:2008bk}. 
UF takes place also in  Bowman's capsule of the human kidneys where water 
and other small molecules are separated from blood \cite{Grimellec:1975}.

The term UF refers to the membrane filtration of smaller, 
submicron-sized particles using larger trans-membrane pressure (TMP) values. It should be distinguished from the so-called microfiltration (MF) of suspensions of larger, micron-sized particles, operated at relatively low TMP values and high permeation fluxes \cite{Belfort:1993if}. While the effect of thermal Brownian motion is strong in UF, causing the dispersion to remain basically in local thermodynamic equilibrium during the filtration process, the dominant particle-dynamics mechanism in MF is shear-induced hydrodynamic migration and non-isotropic collective hydrodynamic diffusion. The important role of the thermodynamic osmotic pressure in UF is taken over in MF by an effective osmotic pressure of hydrodynamic origin which can be characterized by an effective temperature \cite{Yurkovetsky:2008ev,Vollebregt:2010jk}. 

A standard way of operating UF is the inside-out cross-flow mode where a feed dispersion is steadily pumped through a bundle of hollow fiber membranes with inlet and outlet ports. The particle-enriched dispersion is collected at the  outlet ports. Given a fully particle retentive membrane, a small fraction of the in-flowing solvent permeates the membrane to the outside of the fiber into the permeate bath. The particle advection towards the inner membrane wall by the permeating solvent, driven by the TMP, 
is balanced by the diffusive back transport of particles 
away from the inner (lumen-side) membrane wall. 
This leads to the formation of a steady-state concentration-polarization (CP) layer, i.e. a particles-enriched boundary layer of rejected mobile particles near the membrane surface which grows in concentration and thickness with increasing axial distance from the fiber inlet. Owing to the resulting osmotic pressure buildup in the CP layer counteracting the TMP, the permeate flux is decreased and the UF performance is consequently reduced. The increased viscosity in the CP layer causes the flow near the membrane to slow down which in turn further enhances polarization. The flux value above which the flux-TMP curve starts to deviate from the linear pure solvent line is referred to as the critical flux. It hallmarks a detectable influence by the CP layer, or the onset of (irreversible and reversible) membrane fouling mechanisms \cite{Bacchin:2006}.      

For a realistic macroscopic-level calculation of the axially varying CP and permeate velocity profiles, accurate expressions / results are needed for the collective (gradient) diffusion coefficient, $D_c$, and the low-shear dispersion viscosity, $\eta$, in their dependence on the particle volume fraction $\phi$ and the dispersion ionic strength. In addition, accurate expressions for the particle osmotic pressure, $\Pi$, and the related osmotic compressibility, $\chi_\text{osm}$, are required.                     

There exists a larger number of theoretical works on the UF of charge-stabilized dispersions, both for non-steady dead-end (frontal) \cite{Bowen:1995ft,Chun:2001,Trompette:2005hu,Bouchoux:2014kb} and steady cross-flow \cite{Elimelech:1998,Bacchin:2002,Bhattacharjee:1999cz} setups. In most of these works, the concentration and salt content (salinity) dependence of $\Pi$ and $D_c$ has been described using approximate expressions and numerical results. For solutions of proteins such as BSA and lysozyme, virial-expansion type phenomenological expressions for $\Pi$ are frequently used (see, e.g., Refs. \cite{Sarkar:2009,Sarkar:2010du}), while for charge-stabilized colloids, $\Pi$ is approximated by a superposition of hard-sphere, van der Waals and cell model electric pressure contributions \cite{Holt:1995jn,Bowen:1995ft}. The concentration and ionic strength  dependence of $\eta$ is often disregarded in these studies \cite{Bacchin:2002,Bhattacharjee:1999cz}, or it is related to that of $D_c$ by taking for granted (unjustified, as we are going to show) the validity of a generalized Stokes-Einstein relation (cf. Ref. \cite{Jonsson:1996he}). Moreover, the phenomenological Krieger-Dougherty expression for $\eta$ is often used, with the maximum effective particle packing fraction in this expression related to an effective particle diameter quantifying the range of the electric double layer repulsion \cite{Bowen:2007ix}. To describe $D_c$ and its associated 
sedimentation coefficient $K$, cell model expressions are often used \cite{Bacchin:2002} which do not account for particle correlations. These correlations are significantly different in dispersions of charged and neutral particles  
\cite{Banchio:2008gt,Gapinski:2014fn}. 
Only to a small extent have statistical mechanics approaches been used for calculating $K$ and $D_c$, based on an effective particle pair potential combined with an Ornstein-Zernike integral equation method.  
Results obtained in this way are  mainly for high-salinity systems 
where the simplifying concept of an effective hard-sphere diameter can be applied \cite{Bowen:2007ix}, 
and for the solvent-mediated particle hydrodynamic interactions (HIs) being neglected or treated highly approximatively  (see, e.g., Ref. \cite{Bhattacharjee:1999cz}). The inaccuracy of these transport coefficient results is reflected 
in the respective UF model predictions.   

Low-salinity dispersions require special considerations also regarding the osmotic pressure, since the influence of the small counterions dissociated from the particle (i.e., colloid or protein) surfaces and of added salt ions,  termed microions for short, is here crucial. For these systems, the effective particle pair potential with integrated-out microion degrees of freedom is state-dependent, and a simplifying mapping on an effective hard-sphere system is   inadequate for the calculation of static and dynamic properties. 

In this paper, we present and evaluate accurate semi-analytic expressions for thermodynamic, static and dynamic dispersions properties constituting the input to our cross-flow UF model of charge-stabilized globular particle systems under low-salinity conditions. Our calculations of $D_c$ and $\eta$ are based on the one-component macroion-fluid model (OCM) describing microion-dressed particles interacting by an effective pair potential of Derjaguin-Landau-Verwey-Overbeek (DLVO) type. The potential is characterized by a concentration and ionic strength dependent renormalized particle charge number, $Z_\text{eff}$, and an electrostatic  interaction screening parameter $\kappa_\text{eff}$.  To account for the strong effect of surface-released counterions on the OCM potential and osmotic pressure, we use the Poisson-Boltzmann (PB) cell 
model in a form convenient for applications as obtained by Trizac {\em et al.} \cite{Trizac:2003ij}. 
The cell model is combined with the hypernetted chain (HNC) integral equation method of calculating the particle radial distribution function and static structure factor that in turn are used in our calculation of $D_c$ and the high-frequency, $\eta_\infty$, and the shear relaxation, $\Delta\eta$, contributions to the low-shear viscosity $\eta$. 
The semi-analytical methods used here in calculating these transport coefficients have been well assessed in their good accuracy by the comparison with dynamic simulation results where HIs are fully included \cite{Banchio:2008gt,Heinen:2011if}, and with experimental data for charge-stabilized colloidal particles \cite{Westermeier:2012jk} and protein solutions \cite{Gapinski:2005cy,Heinen:2012iy}. The osmotic pressure and transport coefficient results 
constitute the input to the here employed cross-flow UF boundary layer model previously applied by Roa {\em et al.} \cite{Roa:2015dj} to solvent-permeable hard-sphere suspensions. Results are generated and discussed for the two-dimensional CP layer distribution, and for the axially resolved permeate flux.  

Additionally to our theoretical work, we have performed high-precision measurements of the fiber-length averaged permeate velocity and TMP, and the membrane hydraulic resistance, using a well-characterized aqueous suspension of charged silica spheres as the feed. The measurements were made using a specially designed cross-flow filtration device. 
As shown in the paper, 
the theoretical predictions for the permeate flux are in good agreement with those obtained from the filtration experiment.

The essentials of the employed UF boundary layer model are given in Subsec. \ref{sec:UF model}, with the attention turned  on the specifics originating from the particle and membrane charges. 
The cell model calculations of the 
concentration-dependent effective particle charge and screening parameter 
are discussed in Subsec. \ref{sec:ecp}. The osmotic pressure calculations are explained in Subsec.\ref{sec:osmoticpressure}, and the transport coefficients calculations relevant to UF are described in Subsec. \ref{sec:transport}. The details of the cross-flow UF experiments 
on low-salinity aqueous silica particle suspensions are included in Sec. \ref{sec:exp}. 
In Sec. \ref{sec:results}, we present the theoretical predictions for the CP layer and permeate flux profiles, 
and the experimental-theoretical comparison regarding the TMP dependence of the permeate.  
The summary with conclusions is given in Sec. \ref{sec:conclusions}.            
\section{Theoretical ultrafiltration modeling}\label{sec:model}

We describe in this section our macroscopic cross-flow UF modeling for a feed suspension of charge-stabilized colloidal particles, and explain our methods of calculating $D_c$, $\eta$, and the osmotic pressure $\Pi$, based on the OCM with renormalized particle charge and screening parameter. The suspension is steadily pumped through a hollow cylindrical fiber membrane of inner radius $R$, length $L \gg R$, and clean-membrane hydraulic permeability, $L_p^0$, triggered by the applied pressure difference, $\Delta p_L = p_\text{in} - p_\text{out}$, between the inlet and outlet of the fiber. We assume the membrane to be fully retentive 
to the colloidal particles. A small fraction of the axially in-flowing solvent permeates the membrane to the outside of the fiber into the permeate bath. In this inside-out cross-flow setup, the particle advection towards the membrane by the permeating solvent, driven by the difference between the transmembrane pressure, $\Delta p_\text{TMP}$, and transmembrane osmotic particle pressure, $\Pi$, is balanced by the diffusive back transport of particles away from the inner membrane surface. This leads to the formation of a particle-enriched CP region of mobile particles close to the inner membrane surface. A detailed discussion of the physical principles and assumptions underlying diffusion-advection transport in cross-flow UF has been given in \cite{Roa:2015dj}, for suspensions of uncharged, solvent-permeable particles such as non-ionic microgels. Therefore, only the essentials of the diffusion-advection transport are summarized here, with the focus set instead on 
salient features specific to charged-stabilized particles.             

\subsection{Stationary cross-flow transport}\label{sec:UF model} 

In UF, the transport of Brownian particle suspensions is considered under laminar flow conditions where the system is only slightly perturbed from thermal equilibrium. Under continuous cross-flow operation, 
a steady-state is quickly reached, with fully developed suspension flow in the lumen side of the fiber, 
and a particles-enriched stationary CP layer formed at the inner membrane wall. Owing to the very large Schmidt number of colloidal suspensions, given by the ratio of the characteristic single-particle diffusion and hydrodynamic vorticity diffusion times across a distance equal to the colloidal particle radius $a$,  
the steady suspension flow is much faster developed than the CP layer profile \cite{RomeroDavis:1990}. 
The CP layer is more pronounced, and more extended, with axial increasing distance from the fiber inlet. 
On a coarse-grained length scale where the size of the particles is not resolved, the stationary transport is governed by continuum mechanics equations. There is first the mass balance (particle conservation) described by the continuity equation,
\begin{equation}
 \nabla\cdot {\bf J}({\bf r})=0\,,
\label{eq:continuity}
\end{equation}  
where 
\begin{equation}
 {\bf J}({\bf r}) = -D_c(\phi({\bf r})) \nabla \phi({\bf r}) + \phi({\bf r}) {\bf v}({\bf r})
\label{eq:current}
\end{equation}  
is the particle flux, and $\phi({\bf r})$ is the local particle volume fraction at 
position vector ${\bf r}$ inside the suspension. The total flux ${\bf J}({\bf r})$ has a diffusion flux contribution 
related to Brownian motion whose strength at a given local concentration gradient is quantified 
by the so-called collective or gradient diffusion coefficient $D_c(\phi)$, and an advection flux contribution 
proportional to the suspension-averaged fluid velocity ${\bf v}$. 
The momentum balance for the suspension-averaged fluid flow is governed, 
under local low-Reynolds number conditions met in UF, by the effective Stokes equation in conjunction with the incompressibility constraint
\begin{equation}
 \nabla \cdot \bm{\sigma}({\bf r}) = {\bf 0}\,, \qquad \nabla \cdot {\bf v} = 0\,.
\label{eq:Stokes}
\end{equation}  
Here,  
\begin{equation}
 \bm{\sigma} = - p\bm{1} + \eta(\phi)\big[\nabla {\bf v} + \left(\nabla {\bf v}\right)^T \big]
\label{eq:stress}
\end{equation}  
is the suspension-averaged hydrodynamic stress tensor, and ${\bf 1}$ is the unit tensor.   
The superscript $T$ denotes matrix transposition. 
Moreover, $p({\bf r})$ is the suspension-averaged local pressure, 
and $\eta(\phi)$ the effective suspension viscosity for steady low-shear flow. 
According to Eq. (\ref{eq:Stokes}), 
in inhomogeneous suspension regions such as the CP layer, 
there is an additional hydrodynamic force density proportional to $\nabla \eta$. 

The governing Eqs. (\ref{eq:continuity}) - ({\ref{eq:stress}) are subjected to boundary conditions imposing the inlet flow, and specifying the flow conditions at the lumen side of the membrane. 
We assume a fully developed Poiseuille inlet flow,
\begin{equation}
 {\bf v}(r,x=0)= u_m \left(1 - \frac{r^2}{R^2} \right){\bf e}_x,
\label{eq:Poiseuille}  
\end{equation}
of a homogeneous feed solution of (small) volume fraction $\phi_0$. 
The axis of the cylindrical fiber extends from $x=0$ to $L$ into the direction of the unit vector ${\bf e}_x$, 
with $r$ denoting the radial distance from this axis. The inflow velocity, $u_m$, at the fiber axis 
is related to the characteristic shear rate, $\dot{\gamma}$, by \cite{Probstein:2005vy}
\begin{equation}
 \dot{\gamma}= \frac{2 u_m}{R}= \frac{4 Q_\text{feed}}{\rho \pi R^2}\,,
 \label{eq:shearrate}
\end{equation}
where $Q_\text{feed}$ is the integral suspension mass flow through the inlet cross section, and $\rho$ 
is the constant suspension mass density. The mass density difference 
between particles and fluid is neglected here. 

Furthermore, we impose Darcy's law in the integral form \cite{Probstein:2005vy}
\begin{equation}
 v_w(x)\equiv{\bf v}(R,x)\cdot {\bf e}_r = L_p^0 \Big[\Delta\;\!p_\text{TMP}(x) - \Pi\left(\phi_w(x)\right)\Big] \,,
\label{eq:Darcy}
\end{equation}
for the (reverse-osmosis) inside-out permeate velocity, $v_w(x)$, at the membrane surface at axial position $x$. 
Here, $\phi_w(x) = \phi(r=R,x)$ is the particle concentration, $\Delta p_\text{TMP}(x)$ is the transmembrane pressure (TMP),  
and $\Pi(\phi_w)$ is the osmotic pressure at the inner membrane wall, while ${\bf e}_r$ is 
the radial unit vector of the cylindrical coordinate system. 
Moreover, $L_p^0=1/(\eta_0R_\mathrm{mem})$ is the solvent permeability of the clean membrane, and $\eta_0$ is the clean fluid viscosity. 
We use furthermore the zero-tangential fluid velocity condition at the membrane-suspension interface,
\begin{equation}
  {\bf e}_r \times {\bf v}(R,x) \times {\bf e}_r = {\bf 0}\,,
\end{equation}
and the reflecting boundary condition, 
\begin{equation}
  {\bf J}(R,x) \cdot{\bf e}_r =0\,,
\label{eq:noflow}  
\end{equation}
describing the particle-impermeability of the membrane.     

Strictly speaking, the boundary conditions in Eqs. (\ref{eq:Poiseuille}) - (\ref{eq:noflow}) 
should be taken not right at the membrane-suspension interface, but at the external boundary of a transition layer adjacent to the inner membrane surface of thickness $\delta^\ast$, 
which is required to be large compared to the particle size and mean pore size of the membrane, but small compared with the membrane thickness and fiber radius $R$. Furthermore, 
for charged particles and /or a charged membrane, the transition layer thickness $\delta^\ast$ should be large compared to the Debye screening length, 
but small compared to the extension of the CP layer. Provided such a thin transition layer can be 
introduced, and the convective flow driven by the transmembrane pressure does not significantly perturb 
the thermodynamic equilibrium within the layer, Eqs. (\ref{eq:Poiseuille}) - (\ref{eq:noflow}) can be used 
which implicitly imply an infinitely thin transition layer. In particular,  
the filtration behavior is then not affected by the membrane surface charge. 

A more detailed discussion of the transition layer picture, and of near-membrane ion concentration and electric field re-distributions occurring at high filtration rates, will be given in a forthcoming article 
describing a systematic theoretical analysis of the filtration of charge-stabilized suspensions for varying 
salt content and filtration rates conditions. In the present UF study, the requirements for an 
unperturbed transition layer are met for the encountered lower filtration rates. Moreover, 
since $u_m \gg v_w^0$ where $v_w^0 = L_p^0\;\!\Delta\;\!p_\text{TMP}$ is the maximal permeate velocity reached for a clean membrane and pure solvent as the feed,  
also the CP layer is thin compared to $R$. Thus, a boundary layer analysis of Eqs. (\ref{eq:continuity}) - (\ref{eq:stress}) can be made, resulting in a similarity solution for the CP layer concentration profile $\phi(x,y)$, where $y = R-r \ll R$ is the transversal distance from the membrane wall. 
From this profile, and for known concentration dependence of the osmotic pressure, 
the permeate velocity $v_w(x)$ is obtained using Darcy's law in Eq. (\ref{eq:Darcy}).  

The coupled set of non-linear ordinary differential equations from which the 
similarity solution $\phi(x,y)$ is obtained using the boundary conditions noted before, 
is described in detail in \cite{Roa:2015dj} and will thus not be repeated here. 
For given inlet feed 
flow and $\Delta p_L$, the only input required for the 
numerical solution are  
$\Pi(\phi)$, $\eta(\phi)$, and $D_c(\phi)$ characterizing bulk properties 
of the charge-stabilized suspension, 
and $\Delta p_\text{TMP}$. 
Like in Ref. \cite{Roa:2015dj}, 
$\phi(x,y)$ has been calculated using the MATLAB routine bvp4c \cite{Kierzenka:2001bsb}.                      

\subsection{Effective colloid potential}\label{sec:ecp}

As pointed out above, for the theoretical determination of the CP profile and permeate flux the knowledge of the collective diffusion coefficient, $D_c(\phi)$, and steady low-shear viscosity, $\eta(\phi)$, 
of the suspension are required as functions of the colloidal volume fraction $\phi=(4\pi/3)n a^3$, 
with $n$ denoting the colloid number concentration, 
in addition to the suspension osmotic pressure $\Pi(\phi)$ and the related 
osmotic compressibility $\chi_\text{osm}(\phi)$. 
In principle, these properties can be obtained on basis of a so-called Primitive Model (PM) treatment  \cite{Belloni:2000vj,Nagele:1996hr,Hansen:2006uv} where the large multi-valent colloidal particles (termed macroions for short), and  the small surface-released counterions and electrolyte ions (microions), are treated on equal footing as different species of uniformly charged hard spheres immersed in a dielectric structureless Newtonian fluid 
of dielectric constant $\epsilon$ and viscosity $\eta_0$. 
PM-based theoretical calculations and computer simulations are in general quite elaborate, 
owing to the involved different length and time scales characteristic of the spatio-temporal coupling of the different macroion and microion species. 

In taking advantage of the strong size asymmetry of microions and monodisperse assumed colloidal macroions, the one-component macroion fluid model (OCM) is frequently used. In the OCM, the effective pair interaction potential, $u_\text{eff}(r)$, between two microion-dressed charged colloid spheres of radius $a$ at centre-to-centre distance $r$ is modeled, to decent accuracy in general, by the sum of a hard-sphere and screened Coulomb potential of the form \cite{Belloni:2000vj,Nagele:1996hr,Dobnikar:2006ck,Dobnikar:2014hn}
\begin{eqnarray}
\beta u_\text{eff}(r) = l_B Z_\text{eff}^2 \left( \frac{\exp\{\kappa_\text{eff}a\}}{1+\kappa_\text{eff}a} \right)^2
             \frac{ \exp\{-\kappa_\text{eff} r\} }{r}\,, 
\label{eq:OCM potential}
\end{eqnarray}                  
valid for $r > \sigma$ with $\sigma=2a$ denoting the particle diameter.  
Here, $\beta =1/(k_BT)$ is the inverse thermal energy, $l_B=e^2/(\epsilon k_B T)$ is the Bjerrum length of the suspending fluid, and $e$ is the proton charge, while $Z_\text{eff}$ and $\kappa_\text{eff}$ are the, in general,  
concentration and temperature dependent effective colloid charge number and screening parameter, respectively. 
The OCM potential is state dependent as a consequence of having traced out the microion degrees of freedom by starting, e.g., from the multi-component PM description. 
In using the OCM, it is assumed that van der Waals attraction and other non-electric short-range colloid-colloid interactions are negligible. This assumption is justified for charged colloids if the salt concentration is small enough that near-contact configurations are unlikely, 
or if the solvent dielectric constant nearly matches that of the particles, or if the charged particles are additionally sterically stabilized by surface-grafted short polymers \cite{Kitty:2013}. 
Systems describable by the OCM model range from 
charge-stabilized suspensions of rigid colloidal spheres \cite{Westermeier:2012jk} 
to ionic microgels \cite{Holmqvist:2012hv} 
and globular protein solutions \cite{Heinen:2012iy,Gapinski:2005cy}. 
We use here the OCM for calculating the colloid radial distribution function (RDF), $g(r)$,  
and associated colloid static structure factor, $S(q)$,  
that in turn are needed for the calculation of $\eta$, $K$ and $D_c$.   

For monovalent microions that can be treated as pointlike in comparison with the colloidal macroions, and with small microion correlation effects disregarded, $Z_\text{eff}$ and $\kappa_\text{eff}$ can be obtained using the mean-field Poisson-Boltzmann (PB) spherical cell model description of Alexander {\em et al.} \cite{Alexander:1984} (see also {Trizac {\em et al.} \cite{Trizac:2003ij,Dobnikar:2006ck}).
In the PB cell model, the bulk suspension is represented by a single spherical macroion 
with uniformly distributed bare surface charge $Z_\text{bare} e$, 
placed at the centre of a spherical cell  
whose radius $R= a/\phi^{1/3}$ is set by the colloid volume fraction. 
The fluid and microions, with the latter described in the mean-field treatment 
by continuous concentration profiles,  
are confined to the outer shell of thickness $R-a$. 
For a system with monovalent counterions dissociated from the colloid surfaces such as 
for the considered 
silica suspension in osmotic equilibrium with a strong 1-1 electrolyte reservoir 
with concentration $c_\text{res}$ of salt ion pairs, 
the mean-field electrostatic potential, $\Phi(r)$, in units of $k_B T\;\!/e$     
is the solution of the non-linear PB equation \cite{Trizac:2003ij}
\begin{eqnarray}
\Phi''(r) + 2\;\!\frac{\Phi'(r)}{r}= \kappa_\text{res}^2 \sinh\{\Phi(r)\} \,. 
\end{eqnarray}                  
Here, $\kappa_\text{res}^2 = 8\pi l_B c_\text{res}$ is the square of the reservoir electrostatic sceening constant,   
and the prime denotes differentiation with respect to the radial distance $r$. The appropriate inner and outer boundary conditions rendering the solution $\Phi(r)$ unique are $\Phi'(a)=-l_B Z_\text{bare}/a^2$ and $\Phi'(R)=0$, respectively. They express, respectively, that right at the colloid surface there is no electrostatic screening, and that the cell is overall electroneutral.  

Following Alexander {\em et al.} \cite{Alexander:1984}, 
the effective colloid charge number, $Z_\text{eff}$, is then obtained from the solution, 
$\Phi_l(r)$, of the PB equation linearized at the cell boundary 
by using $\Phi'_l(a)=-l_B Z_\text{eff}/a^2$. This leads to \cite{Trizac:2003ij}
\begin{eqnarray}
  \frac{l_B}{a}\;\!Z_\text{eff} = \gamma_R\;\!F(\kappa_\text{eff}a,\phi^{-1/3}) \,,
\end{eqnarray}                  
and 
\begin{eqnarray}
  \kappa_\text{eff}^2 = 4\pi l_B \big[n_{+}(R) + n_{-}(R)\big] = \kappa_\text{res}^2 \cosh\{\Phi(R)\} \,,
\end{eqnarray}                  
with 
\begin{align}
  F(x,y) = \frac{1}{x}\big[&\left(x^2 y-1\right)\sinh\{x \left(y -1\right)\} \nonumber \\
  &+ \;\;x\left(y-1\right)\cosh\{x\left(y-1\right)\}\big] \,,
\end{align}                  
and $\gamma_R=\tanh\{\Phi(R)\}$. Here, $n_\pm(R)$ are the co- and counterion concentrations at the cell boundary, and $\Phi(R)$ is denoted as the Donnan potential. Note that the constant value of the reservoir electrostatic potential is taken here to be zero. 
The concentration $n_s=N_s/V_R$   
(with $V_R=(4\pi/3)R^3$) of co-ions (salt-ion pairs) in the suspension is in general smaller than the reservoir concentration $c_\text{res}$. It is obtained from solving first numerically  
the non-linear PB boundary value problem for the total reduced potential profile $\Phi(r)$, for which we use the MATLAB routine bvp4c. Using this profile, $n_s$ follows then from the volume average of the co-ion profile, $n_{-}(r) =c_s \exp\{\Phi(r)\}$, according to (with $x=r/a$)
\begin{eqnarray}\label{eq:ns-vs-cs}
 \frac{n_s}{c_\text{res}} = 3\;\!\phi \int_1^{R/a} dx\;\!x^2 \exp\{\Phi(x)\}\,, 
\end{eqnarray}                  
where $Z_\text{bare}<0$ and thus $\Phi(x)<0$ have been used. In using the PB cell model to obtain $Z_\text{eff}$ and $\kappa_\text{eff}$ as functions of $(l_B/a)Z_\text{bare}$, $\kappa_\text{res} a$ and $\phi$, we ignore chemical charge regulation effects arising from an incomplete dissociation of colloidal surface ion groups. 
\begin{figure}[t!]
\begin{center}
\includegraphics[width=1\linewidth]{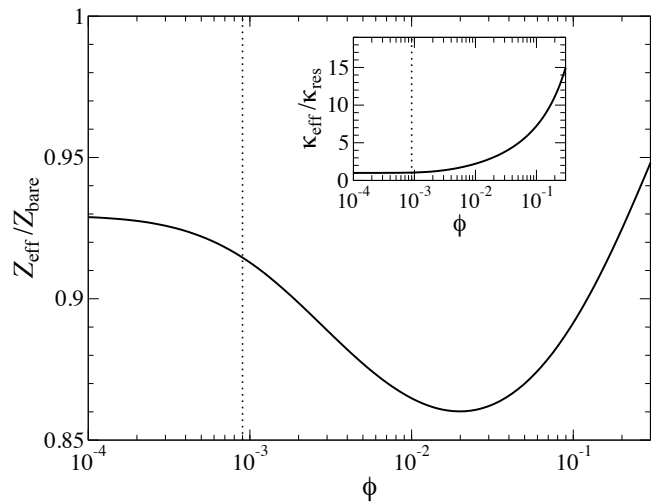}
\caption{Charge number ratio, $Z_\text{eff}/Z_\text{bare}$, (main figure part) 
and screening parameter ratio, $\kappa_\text{eff}/\kappa_\text{res}$, (inset) 
as functions of colloid volume fraction $\phi$, calculated using the Alexander PB cell model. 
The employed system parameters $(l_\mathrm{B}/a)Z_\mathrm{bare}=5, \kappa_\mathrm{res}a=0.15$, $a=15$ nm, and $l_B=0.71$ nm are those of the experimentally studied low-salinity aqueous silica particles suspension in osmotic equilibrium with an aqueous 1-1 electrolyte reservoir. The vertical dotted line marks the threshold concentration value $\phi^\ast$. See the text for details.}
\label{fig1}
\end{center}
\end{figure}

In our silica filtration experiments, the role of the microion reservoir is played by the permeate. On assuming that the membrane is fully retentive to the silica spheres, the ion concentration in the aqueous reservoir is practically set by the water-adsorbed atmospheric $\text{CO}_2$, resulting in $\kappa_\text{res} a = 0.15$. The Donnan equilibrium  corresponds to a semi-grand canonical ensemble description where microions can be freely exchanged between suspension and reservoir. A closed suspension of given salt concentration $n_s$ is treated most easily by mapping it on a corresponding semi-open system. The reservoir concentration $c_\text{res}$, for which $c_\text{res}>n_s$,  
is then uniquely determined from solving Eq. (\ref{eq:ns-vs-cs}), for given $n_s$, using a root-finding procedure. 

The PB cell model predictions for the concentration dependence of the effective charge number and screening parameter of the low-salinity aqueous silica suspension in Donnan equilibrium are depicted in Fig. \ref{fig1} and its inset, respectively, for a concentration independent bare charge number $Z_\text{bare} = 106$, in units of the elementary charge. Owing to the quasi-condensation of counterions at the colloid spheres surfaces, 
the effective charge number $Z_\text{eff}$ is in general smaller than $Z_\text{bare}$. 
For very small $\phi$, the salt ions contribute dominantly to the electrostatic screening (salt-dominated regime), 
and $Z_\text{eff}$ and $\kappa_\text{eff}$ are nearly concentration independent. This is the regime where the OCM effective potential $u_\text{eff}(r)$ is practically state-independent, with values of $\kappa_\text{eff}$ close to the reservoir value $\kappa_\text{res}$ constituting a lower bound. At sufficiently large $\phi$, screening is mainly due to the non-condensed part of the surface-released counterions (counterion-dominated regime). In this higher concentration regime, both $Z_\text{eff}$ and $\kappa_\text{eff}$ change significantly with increasing $\phi$, giving rise to a distinctly state-dependent OCM potential, and values of $\kappa_\text{eff}$ significantly larger than $\kappa_\text{res}$. According to Dobnikar {\em et al.} \cite{Dobnikar:2006ck}, the crossover region connecting the two  regimes in the PB cell model is roughly characterized by the threshold concentration value, 
\begin{equation}
\phi^\ast=0.2\times\frac{(\kappa_\mathrm{res}a)^2}{(l_\mathrm{B}/a)Z_\mathrm{bare}}\,,
\end{equation}
which for our low-salinity silica system amounts to $\phi^\ast = 0.9\times 10^{-3}$, 
indicated by the dotted vertical lines in Fig. \ref{fig1}. 
This value is one order in magnitude smaller 
than the concentration $\phi \approx 0.02$ where the minimal (i.e., maximally charge-renormalized) value  
of $Z_\text{eff}(\phi)$ occurs which is $14\;\!\%$ smaller in magnitude than $Z_\text{bare}$. 
In our filtration experiments using silica suspensions, 
the feed concentration is set to $\phi_0=0.001$, giving silica concentration values, $\phi_w(x)$, at the inner membrane wall that are larger than $\phi^\ast$. It is noticed from Fig. \ref{fig1} that the concentration values  
encountered in the UF experiment are part of the counterion-dominated region 
where $Z_\text{eff}$, and hence $u_\text{eff}(r)$, change significantly 
when $n$ is varied.

The OCM potential with its parameters $Z_\text{eff}$ and $\kappa_\text{eff}$ determined by 
the PB cell model, 
is used in our calculation of the colloid-colloid radial distribution function $g(r)$, 
where $r$ is the center-to-center inter-particle distance, and of the related static structure factor,
\begin{equation}
 S(q) = 1 + 4\pi n \int_0^\infty dr\;\!r^2\left[g(r)-1\right]\frac{\sin(qr)}{qr}\,,
\end{equation}
determined in a scattering experiment, with $q$ denoting the scattering wavenumber. 
For numerical simplicity, we have calculated $g(r)$ and $S(q)$ using the hypernetted-chain (HNC) 
integral equation scheme \cite{Hansen:2006uv}. 
While the HNC lacks thermodynamic self-consistency different, e.g.,  
from the more elaborate Rogers-Young scheme (RY) \cite{Rogers:1984ha}, 
it is decently accurate for the here considered lower-salinity systems 
(cf. Heinen {\em et al.} in Ref. \cite{Heinen:2011if}).

\subsection{Osmotic pressure calculation}\label{sec:osmoticpressure}

The total suspension pressure, $P$, caused by the microions and macroions can be formally split \cite{Dobnikar:2006ck,Trizac:2007eq},
\begin{eqnarray}
 P = P_\text{micro} + P_\text{corr}\,, 
\end{eqnarray}\label{eq:Ptotal}                  
into a microion pressure part, $P_\text{micro}$, deriving from the so-called free volume contribution to the total PM free energy and originating from the non-condensed microions, and the correlation pressure part, $P_\text{corr}$, due to  correlations among the microion-dressed colloids. In the considered Donnan equilibrium 
with a low-concentrated monovalent ions reservoir, the osmotic pressure $\Pi$, i.e. the difference between suspension pressure and reservoir pressure, $P_\text{res}$, is given by
\begin{eqnarray}
 \Pi= P-2\;\!c_\text{res}\;\!k_BT\,. 
\end{eqnarray}\label{eq:Posmotic}                  
Consistent with the PB mean-field level of description, $P_\text{res}$ 
is approximated here by its ideal gas form. 
This simplification is justified, since the leading non-ideal (limiting-law) contribution, 
$-k_BT \kappa_\text{res}^3/(24\pi)$, to the reservoir pressure is,  
for $\kappa_\text{res}a=0.15$, 
three orders of magnitude smaller than the ideal gas part .  

For lower-salinity systems of colloids having many surface charges, the microion (counterion) pressure contribution is dominant so that $P \approx P_\text{micro}$ \cite{Dobnikar:2006ck}. 
This holds in particular in our filtration experiments where $\phi > \phi^\ast$   
as discussed below. In the cell model, $P_\text{micro}$ 
is determined by the microion densities at the cell boundary,
\begin{eqnarray}
 \beta P_\text{micro} = n_{+}(R) +n_{-}(R) = 2\;\! c_\text{res}
 \left(\frac{\kappa_\text{eff}}{\kappa_\text{res}}\right)^2 \,, 
\label{eq:Pmicro}
\end{eqnarray}                  
where the second equality holds in PB approximation. 

The correlation pressure part, $P_\text{corr}$, is in general quite different from the pressure, 
$P_\text{OCM}$, obtained from treating the suspension as an effective one-component fluid of dressed macroions 
with the concentration-dependence of the OCM potential disregarded. 
Under isothermal conditions, 
and without significant effective three-body correlation contributions 
coming into play for very low salinity only, 
the total suspension pressure 
can be determined from the generalized virial pressure equation \cite{Hansen:2006uv},
\begin{align}
 \frac{\beta P}{n} - n\;\!\frac{\partial\beta A_0}{\partial n} = 1 &+ 4\pi \phi 
 g(\sigma^+) 
  \nonumber \\
 &- \frac{2\pi}{3}n \int_{\sigma^{+}}^\infty dr\;\!r^3\;\!g(r) \frac{\partial\beta u_\text{eff}(r)}{\partial r}
  \nonumber \\
 &+ 2\pi n^2 \int_{\sigma^{+}}^\infty dr\;\!r^2\;\!g(r) \frac{\partial\beta u_\text{eff}(r)}{\partial n}\,, \label{eq:P-virial-extended} 
\end{align}                 
for a one-component fluid system with concentration-dependent effective pair potential. 
Additionally to the colloidal ideal gas contribution, and contributions associated with the macroion $g(r)$ of contact value $g(\sigma^+)$ on the right-hand-side of Eq. (\ref{eq:P-virial-extended}), 
there is a pressure contribution deriving from the volume free energy, $A_0(n)$, 
whose colloid concentration dependence is basically due to non-condensed counterions owing to the total electroneutrallity constraint. 
While $A_0$ has no influence on $g(r)$ which is determined solely by $u_\text{eff}(r)$, 
it must be accounted for at smaller salinity in order to properly deduce  
thermodynamic properties using an effective one-component treatment. 

The general applicability of Eq. \eqref{eq:P-virial-extended} has been questioned in the literature \cite{Louis:2002tj,Belloni:2000vj}. 
However, at least in the linear screening case of weakly charged colloids where $Z_\text{eff}=Z_\text{bare}$, 
it exactly reproduces the PM pressure, provided a consistent expression for the free volume pressure contribution on the left-hand-side of Eq. (\ref{eq:P-virial-extended}) is used (see Refs. \cite{Chan:2001id,Chan:2001hx,Denton:2010da,Denton:2007jj,Denton:2014um}). The bare OCM pressure, $P_\text{OCM}$,  
in units of $k_BT$ is given by the right-hand-side of Eq. (\ref{eq:P-virial-extended}), however with 
the negative-valued pressure contribution from the concentration derivative of $u_\text{eff}(r)$ being omitted. 
It is a good approximation of the total suspension pressure $P$ for very large salinity values only,  
when the effect of the surface-released counterions is negligible so that 
$A_0$ and $u_\text{eff}(r)$ become $n$-independent. 
%
\begin{figure}[t!]
\begin{center}
\includegraphics[width=1\linewidth]{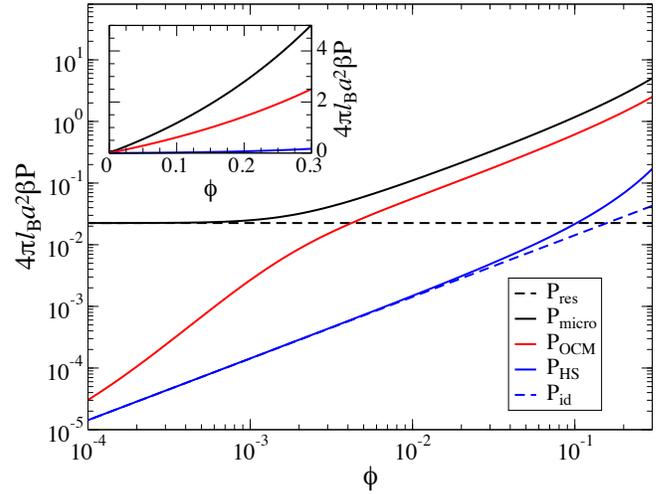}
\caption{Colloid concentration dependence of various reduced pressure contributions listed in the legend, 
for system parameters as in Fig. \ref{fig1}. $P_\text{res}$: ideal gas reservoir pressure; $P_\text{micro}$: PB cell model microionic pressure part according to Eq. (\ref{eq:Pmicro}); $P_\text{OCM}$: OCM pressure part in Eq. (\ref{eq:P-virial-extended}), calculated using the HNC colloid $g(r)$. $P_\text{HS}$: hard-sphere pressure according to Carnahan-Starling equation of state. $P_\text{id}=n k_B T$: colloidal ideal gas pressure. The inset depicts $P_\text{micro}$, $P_\text{OCM}$ and $P_\text{HS}$ on a linear scale.}   
\label{fig2}
\end{center}
\end{figure}

The various pressure contributions are shown in Fig. \ref{fig2}, 
for concentrations extending up to $\phi=0.3$, 
and for the same low-salinity silica system parameters as in Fig. \ref{fig1}.   
Since $g(\sigma^{+})\approx 0$ according to the inset in Fig. \ref{fig3} even for $\phi =0.3$, 
the contact-value 
pressure contribution in Eq. (\ref{eq:P-virial-extended}) is negligibly small. 
Moreover, since the HNC principal structure factor peak height, $S(q_m;\phi)$, 
at wavenumber $q_m$ is   
smaller than $3.1$ for $\phi \leq 0.3$, the suspension  
is liquid-like structured. We have used here the semi-empirical Hansen-Verlet rule stating 
that if $S(q_m) \approx 3.1$ is observed in a charge-stabilized systems with   
$g(\sigma^+) \approx 0$, it is about to crystallize \cite{Gapinski:2014fn}.
%
\begin{figure}[t!]
\begin{center}
\includegraphics[width=1\linewidth]{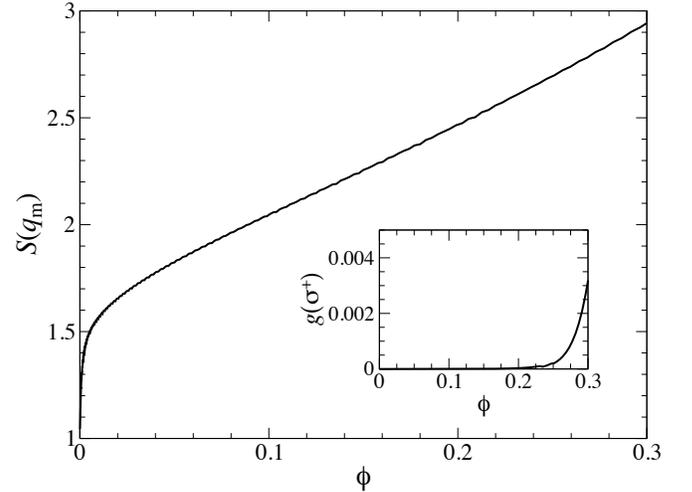}
\caption{Colloid structure factor principal peak height, $S(q_m)$, as function of $\phi$, calculated in HNC approximation using the OCM potential with PB cell model values for $Z_\text{eff}$ and $\kappa_\text{eff}$. System parameters are as in Fig. \ref{fig1}. Inset: Contact value, $g(\sigma^+)$, of the HNC colloid $g(r)$.}   
\label{fig3}
\end{center}
\end{figure}

According to Fig. \ref{fig2}, the main contribution to the suspension pressure arises from the microions. 
That $P \approx P_\text{micro}$ is valid for the silica suspension is an expected feature of systems being part of the counter-ion dominated concentration regime, 
and has been scrutinized in numerous Monte-Carlo simulation studies of 
strongly charge- and size asymmetric PM systems (see, e.g., Refs. \cite{Dobnikar:2006ck,Dobnikar:2014hn}). 
Note further from the figure 
that $P_\text{micro} \gg P_\text{id}$. As a reference, also the pressure curve of a hard-sphere suspension 
is shown in the figure, obtained using the accurate Carnahan-Starling equation of state \cite{Hansen:2006uv}. 
The pressure $P_\text{OCM}$ in Fig. \ref{fig2} is about one half of $P_\text{micro}$, illustrating 
that it strongly overestimates $P_\text{corr}$ in the low-salinity regime \cite{Dobnikar:2006ck,Trizac:2007eq}. 

The key point of the present discussion is that we can use
\begin{equation}
 \beta \Pi \approx \frac{1}{4\pi l_B}\left(\kappa_\text{eff}^2 - \kappa_\text{res}^2 \right)
\label{eq:osmotic-pressure}
\end{equation}
as a good approximation for the osmotic pressure of the low-salinity silica suspension.

In addition to $\Pi$, the long-wavelength limit, $S(0)=S(q \to 0)$, of the macroion 
structure factor is required as another thermodynamic input to the UF model 
contributing to the calculation of $D_c$. 
According to an exact relation by Kirkwood and Buff, 
the isothermal compressibility factor in Donnan equilibrium,  
$\chi_\text{osm}$, can be expressed solely in terms of $S(0)$ as \cite{Kirkwood:1951br} 
\begin{equation}
  \chi_\text{osm}^{-1} \equiv 
  \left(\frac{\partial \beta \Pi}{\partial n} \right)_{T,\text{res}} = \frac{1}{S(0)} \,,
\label{eq:KirkwoodBuff} 
\end{equation}
without an explicit invocation of colloid-microion and microion-microion static correlation functions.  
The concentration derivative of $\Pi$ is taken here for fixed reservoir properties, namely fixed electrolyte ion chemical potential and concentration $c_\text{res}$. Since $P \approx P_\text{micro}$ holds  
for our slow-salinity system, the compressibility factor follows straightforwardly from
\begin{equation}
  \chi_\text{osm}^{-1} \approx \frac{\kappa_\text{eff}}{2\pi l_B} 
  \left(\frac{\partial \kappa_\text{eff}}{\partial n}\right)_{T,\text{res}}\,,  
\label{eq:compressibility}  
\end{equation}
and the PM cell model result for $\kappa_\text{eff}$. 
At low salinity, $\chi_\text{osm} \sim 1/Z_\text{eff}$ is valid approximately,  
rendering explicit  
the low osmotic compressibility of the strongly repelling macroions. 
The Kirkwood-Buff relation is helpful also for testing the degree of self-consistency of the approximations, 
namely here the PB cell model and HNC approximations used  
in the calculations of $\chi_\text{osm}$ and $S(0)$, respectively. 
In the concentration range of our UF study, 
the difference between both calculated quantities is less than 20 \%.   

While the cell-model based $P_\text{micro}$ 
is a good approximation for the suspension pressure of our silica system,  
with compressibility factor determined using Eq. (\ref{eq:compressibility}), 
for completeness and future applications we shortly address how $\Pi$ can be calculated for conditions 
where $P$ is not well approximated any more by $P_\text{micro}$. 
First, a so-called extrapolated-point-charge method of calculating 
$P$ has been developed recently by Boon {\em et al.} \cite{Boon:2015ew}. 
This method invokes likewise the Alexander cell model input for $\kappa_\text{eff}$,  
but now with a different definition of $Z_\text{eff}$ used in the OCM potential.  
Second, in two closely related approaches put forward by Casta\~neda-Priego {\em et al.} \cite{CastanedaPriego:2012fja,CastanedaPriego:2006he}, and Colla {\em et al.} \cite{Colla:2009cz} (see also Ref. \cite{Colla:2012cu}),  
the exact validity of the Kirkwood-Buff relation is enforced by a self-consistent combination of the PB-based renormalized jellium model for calculating $Z_\text{eff}$ and $\kappa_\text{eff}$, 
and the Rogers-Young integral equation scheme for $S(q)$ with adjusted mixing parameter.  
Which of the above noted methods of calculating $\Pi$ is more accurate    
in comparison with benchmark PM simulations is a matter of future assessment. 

According to a multi-colloid PB Brownian dynamics simulation study by Hallez {\em et al.} \cite{Hallez:2014}, the identification of the suspension osmotic pressure $\Pi$ with the microion osmotic pressure calculated using the PB cell model is quite accurate for $\kappa_\text{res}\;\!d \lesssim 1$, but not reliable for $\kappa_\text{res}\;\!d \gtrsim 5$, with a larger error introduced for intermediate values of $\kappa_\text{res} d$ that is strongly depending on $(l_B/a)|Z_\text{bare}|$. Here, $d=a\left(\phi^{-1/3} -1\right)$ is roughly twice the average inter-colloid distance. Denton in Ref. \cite{Denton:2010da} has arrived earlier at a similar conclusion regarding the worsening of the PB cell model in its osmotic pressure prediction for intermediate salt concentrations. For the colloid concentrations in the CP layer of the silica suspension shown in Fig. \ref{fig6}, one obtains $\kappa_\text{res} d$ in between $0.9-1.35$. The error introduced in approximating $\Pi$ by the cell model $\Pi_\text{micro}$ is thus less than 5\% according to Fig. 4 in \cite{Hallez:2014}.  

We finally point to the variational method (free energy minimalization) by Denton \cite{Denton2008,Lu2010,Denton:2010da}. It employs yet another definition of the effective colloid charge and screening parameter, and it accounts for all the concentration-dependent contributions to the osmotic pressure in Eq. (\ref{eq:P-virial-extended}), including the volume free energy contribution. The Denton method is applicable for arbitrary salinity, and its predictions for the osmotic pressure and radial distribution function agree well with PM based Monte-Carlo simulation results by Linse \cite{Linse2000}.

\subsection{Colloidal transport coefficients}\label{sec:transport}

The concentration-dependent collective diffusion coefficient, $D_c(\phi)$, 
in the constitutive equation invoking the coarse-grained silica particles flux ${\bf J}({\bf r},t)$, 
can be expressed in Donnan equilibrium as \cite{Banchio:2008gt}
\begin{equation}
 D_c(\phi) = D_0 \frac{K(\phi)}{\chi_\text{osm}}\,,
\end{equation} 
where $D_0 = k_B T/(6\pi\eta_0 a)$ is the single-particle diffusion coefficient, and $\chi_\text{osm}=S(0)$ is the osmotic compressibility coefficient calculated using Eq. (\ref{eq:compressibility}). 
Here, $K(\phi)=V_\text{sed}(\phi)/V_0$ is the long-time sedimentation coefficient, with $V_\text{sed}(\phi)$ denoting the mean particle sedimentation velocity in a uniform weak force field that reduces to the single-particle sedimentation velocity, $V_0$, at infinite dilution. As discussed in Ref. \cite{Banchio:2008gt}, $V_\text{sed}$ is 
in principle smaller than the  corresponding short-time sedimentation coefficient. 
However, for low-salinity system where two-body hydrodynamic interactions (HIs) are prevailing, the difference between the two coefficients is minuscule and can be ignored. 
Consequently, we can identify $K$ according to Ref. \cite{Heinen:2011if} with
\begin{equation}
  K = H(q\to 0;\phi)\,,
\end{equation} 
i.e. with the zero-wavenumber limit of the so-called colloidal hydrodynamic function $H(q;\phi)$. The latter is routinely determined for colloidal suspensions using 
short-time dynamic light scattering experiments \cite{Holmqvist:2010ki,Riest:2015jo}. 
The function $H(q;\phi)$ contains information about short-time diffusion processes on length scales $\sim 1/q$, 
and for correlation times $t \ll a^2/D_0$ \cite{Nagele:1996hr}. 
For the hypothetical case of hydrodynamically non-interacting particles, $H(q;\phi) =1$ independent of $q$ and $\phi$. Values of $K$ smaller than one are thus a hallmark of the slowing influence of the HIs. 

The short-time function $H(q)$ can be expressed by an equilibrium average invoking a specific combination of hydrodynamic mobility tensors characterizing HIs under low-Reynolds-number flow conditions \cite{Nagele:1996hr}. 
To calculate $H(q)$ from this average, we use the well-established analytic BM-PA method \cite{Heinen:2011if,Riest:2015jo,Beenakker:1983fh,Beenakker:1984hda}. 
This method is a hybrid of the Beenakker-Mazur method (BM), 
used here for the wavenumber-dependent distinct part of $H(q)$, 
and the hydrodynamic pairwise-additivity method (PA) used for the $q$-independent self-diffusion part. 
The BM-PA scheme requires the colloidal $S(q)$ and $g(r)$ as its only input, 
for which the HNC results based on Eq. (\ref{eq:OCM potential}) for $u_\text{eff}(r)$ are used. 
For a charge-stabilized suspension whose colloidal interactions are described by an OCM-type potential, 
the BM-PA method predicts $H(q)$ in good overall agreement with simulation and experimental result \cite{Heinen:2011if,Riest:2015jo,Banchio:2008gt,Westermeier:2012jk}.

\begin{figure}[t!]
\begin{center}
\includegraphics[width=1\linewidth]{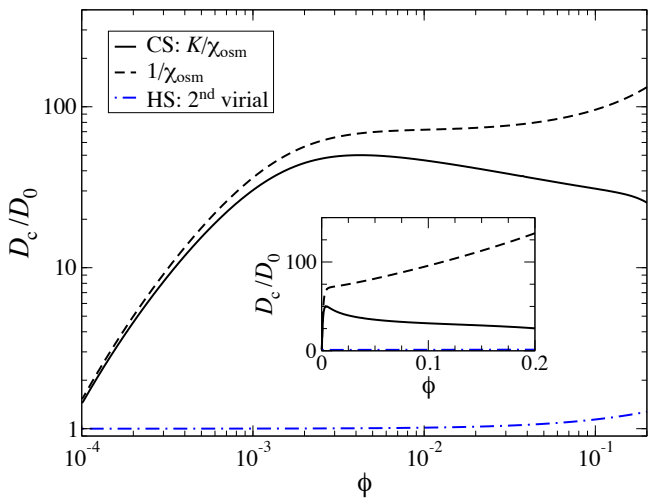}
\caption{Reduced collective diffusion coefficient, $D_c(\phi)/D_0$, as a function of $\phi$, for system parameters as in Fig. \ref{fig1}. Solid line: Charged silica (CS) suspension, with $K=H(0)$ calculated by the BM-PA method using the HNC input for $S(q)$ and $g(r)$, and $\chi_\text{osm}$ calculated using Eq. (\ref{eq:compressibility}). Dashed line: Hypothetical silica suspension without HIs for which $K=1$. Dashed-dotted line: Second-order virial expansion result for neutral hard spheres (HS).}
\label{fig4}
\end{center}
\end{figure}

Fig. \ref{fig4} depicts the concentration dependence of $D_c$ for the silica system, 
calculated using the BM-PA method. This result is compared with the corresponding result
in which the long-ranged colloidal HIs are disregarded, and with the $D_c$ calculation 
for neutral hard spheres. Both the sedimentation coefficient and the osmotic compressibility factor are monotonically decreasing with increasing $\phi$. 
At small $\phi$, the decline 
of $\chi_\text{osm}$ outbalances that of $K$, owing to the strong electrostatic inter-particle repulsion, 
with the consequence that $D_c$ rises steeply initially (see the inset). For a thermally induced concentration fluctuation where the colloid concentration in a small region is larger than in a neighboring one, the relaxation of the local volume concentration gradient, $\nabla \phi({\bf r})$, by the diffusive collective motion of particles described by Fick's law, ${\bf J}_D = -D_c \nabla \phi$ (see Eq. (\ref{eq:current})), is enhanced by the  osmotic pressure difference between the two regions. For larger $\phi$, the slowing influence of the HIs becomes stronger, with the consequence that $D_c$ passes through a maximum at $\phi \approx 0.004$, 
followed by a moderate decline of $D_c$ for larger $\phi$ values. When the HIs are neglected so that $K=1$, 
a monotonically increasing $D_c$ is obtained instead in the considered concentration range. 
The key fact to notice from Fig. \ref{fig4} 
is that owing to the strong electrostatic repulsion between the silica particles, 
$D_c$ is strongly enhanced by one order in magnitude relative to the collective diffusion coefficient 
of neutral hard spheres.  
The hard-sphere result for $D_c$ with HIs included shown in the figure has been generated 
using the second-order virial expansion expression \cite{Cichocki:2002ft,Cichocki:1999da,Banchio:2008gt},
\begin{equation}
 D_c(\phi)/D_0 = 1 + 1.454 \phi - 0.45 \phi^2 \,,
\end{equation}           
which is in good agreement with simulation data up to $\phi = 0.494$ 
where a non-sheared hard-sphere suspension starts to solidify (see, eg., Ref. \cite{Roa:2015dj}).  

Note that an electrokinetic reduction of $D_c$ arising from the non-instantaneous relaxation of the microion clouds surrounding each colloidal macroion, is not accounted for in the BM-PA method based on the OCM. 
This reduction can be estimated using the PM-based coupled-mode theory \cite{Gapinski:2005cy} 
predicting it to be negligibly small 
owing to the large silica-microion size asymmetry.

The second important transport property input to our filtration model 
is the concentration-dependent effective suspension viscosity $\eta(\phi)$. 
Just like $D_c(\phi)$, it depends on the suspension salinity and colloid  
surface charge. The steady-shear viscosity, $\eta$, is the sum \cite{Russel:1984iu},
\begin{equation}
  \eta(\phi) = \eta_\infty(\phi) +\Delta\eta(\phi)\,,
\end{equation}  
of the high-frequency viscosity contribution, $\eta_\infty(\phi)$, of purely hydrodynamic origin, 
and the shear relaxation viscosity contribution, $\Delta\eta(\phi)$, 
due to dissipation originating from the relaxation of the shear-perturbed particle cages formed around each colloidal particle. The viscosity part $\Delta\eta$ is influenced both by  direct and hydrodynamic interactions, with the consequence that in suspensions of strongly correlated particles such as the present one,  
the long-time viscosity $\eta$ is significantly larger than its short-time cousin $\eta_\infty$. 
This distinguishes the viscosity from the collective diffusion coefficient, 
since for the latter 
the difference between its short- and long-time forms stays small. 

Shear-thinning effects can be neglected under UF conditions where the shear Peclet-number is small. 
Moreover, since for moderate salinity near-contact configurations of three or more silica spheres are unlikely, we can use the PA method for calculating the low-shear $\eta_\infty$. 
In this method, two-particle HIs are fully accounted for including near-contact lubrication terms \cite{Riest:2015jo}. 
We have checked that for $\phi \leq 0.3$, 
the PA result for the $\eta_\infty$ of the silica system is quantitatively described by the polynomial
\begin{equation}
 \frac{\eta_\infty}{\eta_0} = 1 + \frac{5}{2} \phi\left(1+\phi\right) + 7.9 \phi^3 \,.
\label{eq:CS-highfrequency}
\end{equation}   
We emphasize that this polynomial is not a truncated virial expansion expression. For low-salinity suspensions with $\kappa_\text{res} a \lesssim 1$, a virial-type expansion of colloidal transport properties based on an effective particle diameter determined, e.g., using the Barker-Henderson perturbation approach has been shown to be not useful (see, e.g., Ref. \cite{Nagele:1996hr}).    
The result in Eq. (\ref{eq:CS-highfrequency}) was derived in Ref. \cite{Banchio:2008gt} using the PA method and additional simplifications justified for low-salinity systems, and it was shown therein to be in good agreement with elaborate hydrodynamic simulation results for $\eta_\infty$. 
Note that $\eta_\infty$ of charged silica particle suspensions 
is somewhat smaller than that of uncharged hard spheres for the same concentration, 
owing to the smaller likelihood of near-contact configurations in the former case. 

While well-tested analytic tools such as the PA method are available for $\eta_\infty$, 
the calculation of $\Delta\eta$ 
is distinctly more demanding since caging (memory) effects need to be considered for the latter. 
There are only few simulation studies on the steady-shear viscosity of charge-stabilized suspensions, 
and in most of them HIs have been neglected. For calculating $\Delta \eta$, 
we employ here the mode-coupling theory (MCT) approximation applied to Brownian particle systems. 
A version of the MCT has been developed by one of the present authors \cite{Nagele:1998in} where for charged particles the most important far-field HIs contributions are included, in conjunction with 
a generalization to the multicomponent PM that was successfully applied to concentrated electrolyte solutions  \cite{Aburto:2013ca}. 
For analytic simplicity, we start here from the standard one-component MCT expression, 
\begin{equation}
  \Delta\eta_\text{MCT} = \frac{k_B T}{60 \pi^2} \int_0^\infty dq\;\!q^4 \left(\frac{S'(q)}{S(q)}\right)^2 
               \int_0^\infty dt\;\!\left(\frac{S(q,t)}{S(q)}\right)^2 \,,
\label{eq:MCT}
\end{equation}
derived, e.g., 
in Ref. \cite{Nagele:1998in}, in which HIs contributions to the static MCT vertex function are neglected.  
The prime denotes here differentiation with respect to the wavenumber. In principle, $\Delta\eta$ can be calculated self-consistently using a numerically expensive algorithm in combination with the corresponding MCT equation for the dynamic structure factor $S(q,t)$, where $S(q,0)=S(q)$ \cite{Nagele:1996hr}. In restricting ourselves to concentration values $\phi < 0.2$ where $S(q_m)$ is distinctly smaller than the Hansen-Verlet freezing criterion value of $3.1$, 
we can obtain $\Delta\eta$ more simply in form of a first iteration solution by replacing $S(q,t)$ in Eq. (\ref{eq:MCT}) by its exponential short-time form $S(q,t)/S(q)=\exp\{-q^2 D_0 H(q) t/S(q)\}$. This gives,
\begin{equation}
  \frac{\Delta\eta_\text{MCT}}{\eta_0} \approx \frac{1}{40\pi} \int_0^\infty dy\;\!y^2\;\!\frac{S'(y)^2}{S(y)}\frac{1}{H(y)}\,,
\label{eq:MCTsimple}
\end{equation}   
with $y=q\sigma$ and the prime denoting differentiation now with respect to $y$. 
The influence of HIs is incorporated here by means of 
the hydrodynamic function $H(y)$ which affects the short-time decay of $S(y,t)$. As shown in Ref. \cite{Nagele:1998in}, HIs modify also the static vertex function part in the MCT expression of $\Delta\eta$. 
This modification is disregarded here since its effect can be expected to be small in the range of smaller concentration values encountered inside the CP layers of our UF experiments (see below).    
 
\begin{figure}[t!]
\begin{center}
\includegraphics[width=1\linewidth]{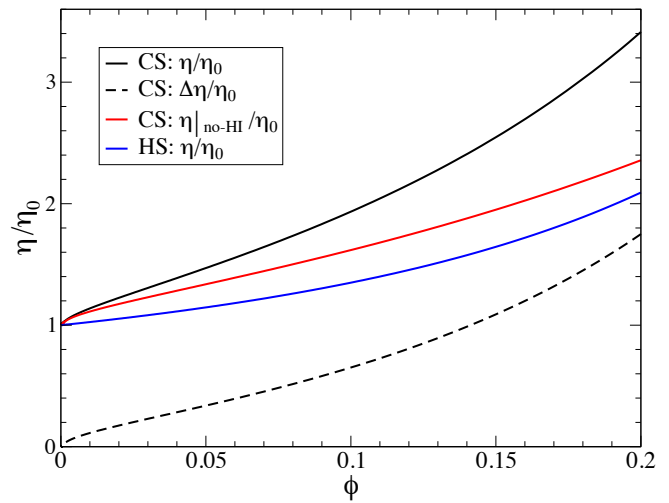}
\caption{Theoretical predictions of the concentration-dependent steady-shear viscosity, $\eta(\phi)$, of charged silica spheres (CS, solid black line), in comparison with the viscosity of neutral hard spheres (HS, solid blue line). Additionally shown are the shear-relaxation viscosity contribution, $\Delta\eta$, of silica spheres, 
and the steady-shear viscosity without HIs. System parameters of the silica system are as in Fig. \ref{fig1}.}
\label{fig5}
\end{center}
\end{figure}

The result for the steady-shear viscosity of the silica system as a function of $\phi$ is included in Fig. \ref{fig5}. The high-frequency viscosity part, $\eta_\infty$, is calculated using Eq. (\ref{eq:CS-highfrequency}), and the shear-relaxation part is obtained using Eq. (\ref{eq:MCTsimple}). We employ here the BM-PA input for $H(y)$, and the HNC $S(y)$ based on the OCM potential with PB cell model values for $Z_\text{eff}$ and $\kappa_\text{eff}$. 
As it is noticed from the comparison of $\Delta\eta$ and $\eta$, 
the high-frequency viscosity and the shear-relaxation part are comparable in magnitude. For $\phi=0.2$, 
$\eta$ is enlarged compared to the solvent viscosity $\eta_0$ by the factor of $3.4$, while 
the high-frequency viscosity is enlarged by the factor of $1.7$. 
Notice here that $\Delta\eta(\phi=0)=0$. 
To quantify the influence of the HIs, in Fig. \ref{fig5} we have included the steady-shear viscosity without HIs, 
for which $\eta_\infty$ is given by the Einstein expression $\eta_\infty/\eta_0 = 1 + 2.5 \phi$ valid for no-slip spheres, 
and for which $H(y)$ in Eq. (\ref{eq:MCTsimple}) is taken to be equal to one.  
HIs significantly enhance the steady-shear viscosity of the silica suspension for $\phi >0.1$.  
For comparison, additionally shown in the figure is the viscosity 
of hard spheres, which we have obtained using a precise generalized Saito formula for $\eta_\infty$, 
and the hard-sphere contact-value approximation for $\Delta\eta$. 
We refer to Refs. \cite{Roa:2015dj,Riest:2015jo} for the details of this analytic calculation, where in addition 
the good accuracy of the generalized Saito formula has been established by the comparison with computer simulation data. 
Different from the charged silica particles,   
the $\Delta\eta$ of neutral hard spheres is for $\phi < 0.2$ small compared to $\eta_\infty$.

Note that we are dealing here with the demanding case of low-salinity systems with 
extended  electric double layers and long-ranged electric repulsion. Electrokinetic viscosity effects 
are not considered here, for these are secondary effects that become smaller 
with increasing concentration. 
In case of higher-salinity systems with thin electric double layers, 
a simple thermodynamic perturbation  
approach is useful where the viscosity calculation can be mapped to that 
for an effective hard-sphere system (see, e.g.,  Ref. \cite{Berli:2005tw}). 

In closing our discussion of transport coefficient calculations, we emphasize that the PB cell model has been used  
only for deriving the effective charge and screening length in $u_\text{eff}(r)$ but not in the calculation of 
the transport coefficients. Thus, different from pure cell model approaches such as that by J\"onsson and J\"onsson \cite{Jonsson:1996he}, colloidal particle correlations are accounted for. 
In \cite{Jonsson:1996he}, the collective friction coefficient $f_c=k_BT/D_c$ 
and thus $D_c$ was estimated on assuming the validity of the generalized Stokes-Einstein (GSE) relation, $D_c/D_0\approx \eta_0/\eta$, between $D_c$ and $\eta$, and by using the spherical cell model viscosity expression. 
This GSE approach is flawed for the following reasons. First, it has been shown  
using theory and simulations, 
and in experimental work on BSA protein solutions \cite{Heinen:2012iy}  
that the aforementioned GSE between $D_c$ and $\eta$ is invalid unless the concentration is very small.  
In fact, according to Fig. \ref{fig4}, the $D_c$ of a lower-salinity system  
has a non-monotonic $\phi$ dependence, 
whereas $\eta$ and $\eta_\infty$ are monotonically increasing with increasing $\phi$.  
Another so-called Kholodenko-Douglas GSE relation between $D_c$ and $\eta$ which in addition invokes the osmotic 
compressibility factor $S(0)$, has been likewise shown to be invalid for low-salinity systems, 
although this relation applies decently well to hard spheres (see Refs. \cite{Heinen:2012iy,Heinen:2011if} for details). 
Second, the comparison with simulation results for $\eta$ and $K$ revealed that cell model predictions for 
these quantities are generally not reliable in particular 
for smaller concentrations \cite{Roa:2015dj}. Moreover, owing to neglected inter-particle correlations, 
the viscosity result obtained from the standard PB cell model scheme is more adequately identified with $\eta_\infty$ rather than with $\eta$. As seen in Fig. \ref{fig5}, 
for a low-salinity system $\eta$ is distinctly larger than $\eta_\infty$.

\section{Silica particles ultrafiltration experiment}\label{sec:exp}

We explain here our cross-flow UF measurements using aqueous suspensions of charge-stabilized silica particles. 
The suspensions consist of Ludox silica particles  
dispersed in purified water without added electrolyte. 
The mean hydrodynamic particle radius which we obtained from dynamic light scattering is $a=15$ nm. The volume fraction of the feed suspension is $\phi_0=1\times10^{-3}$. 
Additionally to monovalent counterions dissociated 
from the silica particles surfaces that neutralize the negative particle charges, 
the suspension includes ions originating from the self-dissociation of water molecules, 
and from atmospheric CO$_2$ contamination. These contributions sum up to the value $\mathrm{pH}=5.5$ that we have measured using a pH meter. 

The number of bare elementary charges on a silica particle surface is estimated as $Z_\mathrm{bare}\approx 106$, i.e. as $(l_B/a)Z_\mathrm{bare}=5$ in reduced units. We have obtained this value using Fig. 4.10 in Ref. \cite{Iler:1979vm} 
from which for $\mathrm{pH}=5$ the surface charge density $0.0375\;\!e/$nm$^{2}$ of SiO$^-$ ions is deduced. 

The Debye screening parameter, $\kappa_\mathrm{res}$, of the permeate reservoir is estimated 
by taking the pH of the permeate to be the same as that of the dilute feed suspension. 
This is justified considering the low silica concentration in the feed, and the fact that microions are not retained by the membrane pores. Moreover, the permeate in the filtration 
device is likewise subjected to CO$_2$ contamination.   
The screening parameter follows then from 
\begin{equation}
\kappa^{2}_\mathrm{res}=8\pi l_B N_A [\mathrm{H}^+]\,,
\end{equation}
where $[\mathrm{H}^+]=10^{-\mathrm{pH}}$ is the molar hydronium concentration, and $N_A$ denotes Avogadro's number. 
This leads to $\kappa_\mathrm{res}a=0.15$. The values for $Z_\text{bare}$ and $\kappa_\text{res}$ 
given above are used in our cell model calculations of $Z_\text{eff}$ and $\kappa_\text{eff}$ 
going into the OCM potential. 
%
%
%
%

We have performed inside-out cross-flow UF measurements using an OSMO Inspector device built by Convergence Industry B.V. (The Netherlands). The setup is equipped with a corioli flow mass flow meter from Bronkhorst Cori-Tech B.V. (The Netherlands) with an accuracy of about 0.2 \%, and the system is based on work by van de Ven {\em et al.} 
\cite{vandeVen:2008}.
In the OSMO inspector device, the silica suspension is steadily pumped, at fixed temperature $T=303$ K, through a membrane module containing 10 hollow cylindrical fiber membranes in parallel mode. 
The employed hollow fiber membrane is a negatively charged polyethersulfon membrane, provided by Pentair X-Flow (The Netherlands), with a nominal pore size of 10 nm and a molecular weight cut-off of 10 kDa.  
The fibers have a mean length $L=40$ cm, and an inner mean diameter $2R=0.8$ mm, 
with total area $A_\mathrm{mem}=2\pi R L=0.01$m$^2$ of the membrane module.

The OSMO Inspector allows to set the feed and retentate (outlet) mass fluxes $Q_\mathrm{feed}$ and $Q_\mathrm{ret}$, respectively. The permeate flux follows from mass conservation as $Q_\mathrm{perm}=Q_\mathrm{feed}-Q_\mathrm{ret}$. Moreover, the mechanical (i.e. non-osmotic) pressure values at the feed, retentate and permeate positions are measured. From these values, the fiber-length-averaged transmembrane pressure, $\Delta p_\mathrm{TMP}$, 
is calculated using
\begin{equation}\label{tmp}
\Delta p_\mathrm{TMP}=\frac{P_\mathrm{feed}+P_\mathrm{ret}}{2}-P_\mathrm{perm}\,,
\end{equation}
where a linear axial pressure drop from inlet to outlet is assumed, 
for a constant permeate pressure. 
Using a linear pressure profile is an approximation sufficient for the present analysis. 
The form of the lumen-side axial pressure decline is actually more complicated, as discussed by Mondor and Moresoli on basis of the momentum and continuity equations combined with Darcy's law \cite{Mondor:1999bd,Mondor:2006kj}.

In our UF experiments, the feed flux was held constant at $Q_\mathrm{feed}=600$ g/h, 
while the retentate flux was stepwise decreased. Accordingly, the permeate flux increased stepwise 
from $Q_\mathrm{perm}=100-500$ g/h. The fiber-length-averaged permeate velocity is obtained using 
\begin{equation}\label{vwav_exp}
\langle v_w\rangle={Q_\mathrm{perm}}/{\rho\;\! A_\mathrm{mem}}\,,
\end{equation}
with the suspension mass density $\rho=1000$ g/L taken to be constant. 
The selected process parameters give rise to measured permeate velocities 
in the range $\langle v_w\rangle=10$ - $50$ LMH  (liters/m$^2$/h), i.e., within $2.8$ -$13.9$ $\mu$m/s.

The clean water permeability, $L_p^0$, of the membrane was measured before and after the silica UF experiment. 
The measurement was done in constant flux mode, with 
$v_w^0=50$ LMH $=13.9\ \mu$m/s, 
and $\Delta p_\mathrm{TMP}$ obtained from inserting 
the associated measured pressure values into Eq. (\ref{tmp}). 
The permeability follows then from Darcy's law, 
\begin{equation}
L_p^{0}=\frac{v_w^0}{\Delta p_\mathrm{TMP}}\,,
\end{equation}
without osmotic pressure contribution since pure water is used as feed. 
In this way, the value $L_p^{0}\approx 155$ LMH/bar$=4.3\times 10^{-10}$ m/Pa$\cdot$s 
was obtained which is used in our UF model calculations of the permeate flux and CP layer profiles discussed in the following section.  

\section{Filtration Results and Discussion}\label{sec:results}

\subsection{Theoretical results for CP layer profiles and permeate flux}\label{sec:restheo}

We present here our theoretical results for cross-flow UF of low-salinity suspensions based on the boundary layer filtration model, and the input for the osmotic pressure and compressibility, and $D_c(\phi)$ and $\eta(\phi)$ as described in Sec. \ref{sec:model}. The system parameters are those characterizing the OSMO Inspector cross-flow UF setup, and the low-salinity aqueous silica suspensions for $T=303$ K. The setup parameters are explicitly: membrane length $L=0.4$ m, inner radius $R=0.4$ mm, hydraulic membrane permeability $L_p^0=4.3\times10^{-10}$ mPa$^{-1}$s$^{-1}$, and characteristic shear rate $\dot{\gamma}=332$ s$^{-1}$ corresponding to the feed flux $Q_\mathrm{feed}=600$ g/h according to Eq. (\ref{eq:shearrate}). For simplicity, in our calculations the TMP is taken as 
constant along the fiber, and with value according to Eq. (\ref{tmp}). 
The silica system parameters used in the filtration calculations are: particle radius $a=15$ nm, $l_B=0.7$ nm, $Z_\text{bare}=106$, and $\kappa_\text{res}a=0.15$. The feed volume fraction $\phi_0=1.0\times 10^{-3}$ is large enough for  Eqs. (\ref{eq:osmotic-pressure}) and (\ref{eq:compressibility}) describing the osmotic pressure and compressibility in the counterion-dominated regime to apply. 
Moreover, $Pe_{\dot{\gamma}}^a \ll 1$ so that shear-induced hydrodynamic diffusion is negligible 
in comparison to thermal diffusion \cite{Belfort:1993if}. Since $R \ll L$, the boundary-layer description 
condition $u_m \gg v_w^0$ is met, and since $Re_R \sim 10$ also the Rayleigh number condition of laminar pipe flow is  fulfilled. We have checked that the (effective) Debye length is much smaller than the (typical) thickness  
of the CP layer. Furthermore, the UF condition of local 
thermodynamic equilibrium is fulfilled.

The CP layer and permeate velocity profiles are calculated using the boundary layer method described in Sec. \ref{sec:model}, and with $D_c(\phi)$, $\eta(\phi)$ and $\Pi(\phi_w)$ used as the input. 
We show in the following that for the present    
operating conditions, membrane fouling due to a cake layer formed by jammed particles is avoided.

\begin{figure}[t!]
\begin{center}
\includegraphics[width=1\linewidth]{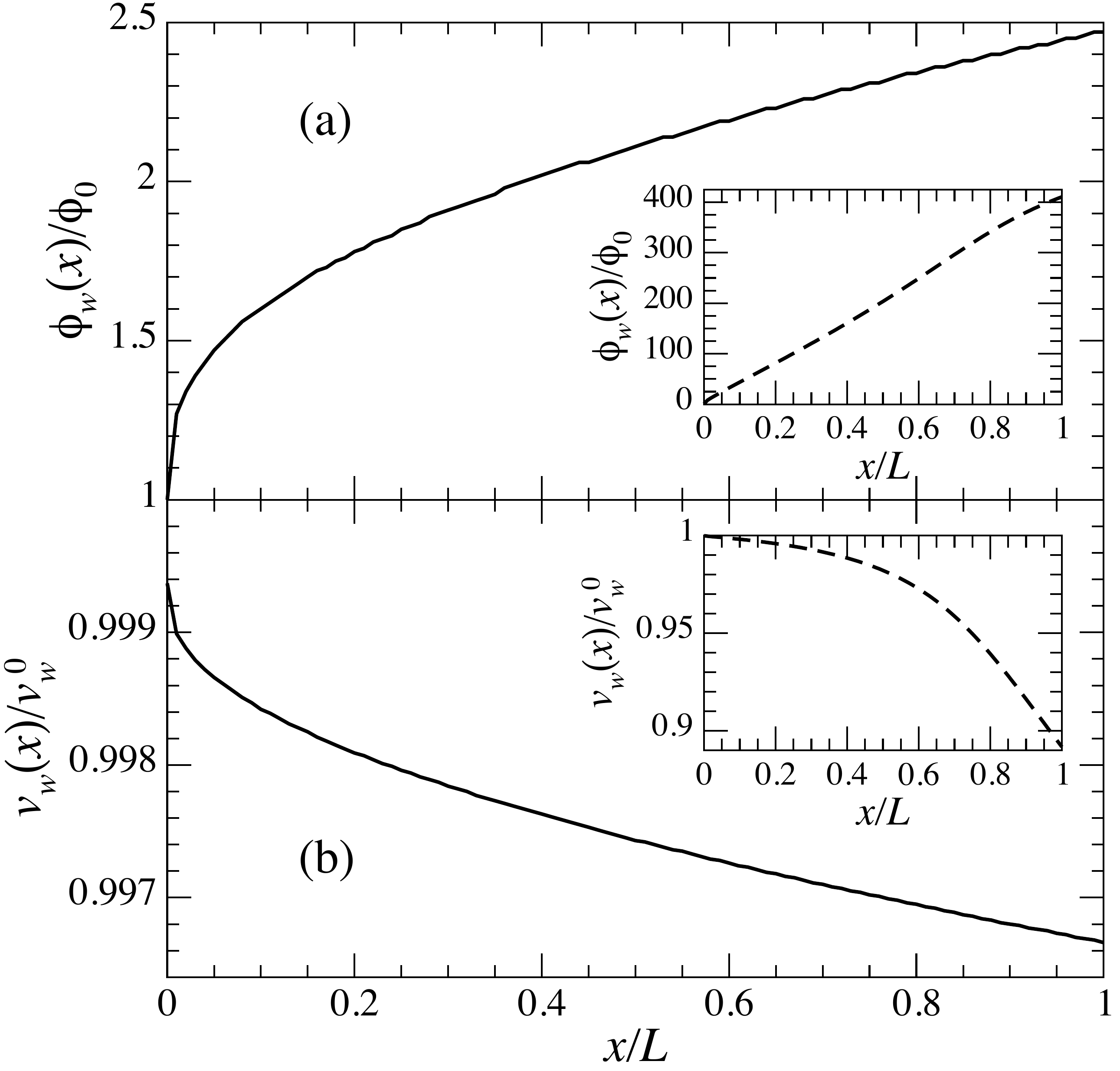}
\caption{(a) Calculated membrane surface concentration profile, $\phi_w(x)$, scaled by the feed volume fraction $\phi_0$, and (b) permeate velocity profile, $v_w(x)$, scaled by the pure solvent velocity, $v_w^0$, 
for the explored low-salinity silica suspension (solid lines). Insets: as in main figures, 
but for neutral hard spheres (dashed lines). System parameters: $\Delta p_\mathrm{TMP}=0.08$ bar, $\dot{\gamma}=332$ s$^{-1}$, $a=15$ nm, $\phi_0=10^{-3}$, $L_p^0=4.3\times 10^{-10}$ m/Pa$\cdot$s.}
\label{fig6}
\end{center}
\end{figure}

Figs. \ref{fig6}(a) and (b) depict the calculated CP concentration profile, $\phi_w(x)$, and the permeate velocity profile, $v_w(x)$, at the membrane surface in their dependence on the reduced axial distance, $x/L$, from the fiber inlet. The solid curves are the results for the low-salinity silica suspension, while the dashed curves in the inset describe neutral hard spheres. Note that for the electrostatically repelling silica particles, $\phi_w(x)$ increases only slightly above the feed concentration $\phi_0$ with increasing distance $x$. This can be attributed to the large values of the collective diffusion coefficient, $D_c(\phi)$, of charge-stabilized particles even for small $\phi$ (see Fig. \ref{fig4}), causing particles which are flow-advected towards the membrane surface to be strongly driven away from it by collective diffusion. In Fig. \ref{fig6}(b), the permeate velocity, $v_w(x)$, of the silica system decreases only slightly with increasing $x$ for the following reason: With increasing axial distance from the inlet, $\phi_w$ and hence $\Pi(\phi_w) \approx P_\text{micro}-P_\text{res}$ (see Fig. \ref{fig2}) are only mildly enlarged so that $v_w(x)$ according to Eq. (\ref{eq:Darcy}) is only slightly lowered below its clear solvent value $v_w^0 = L_p^0\;\!\Delta p_\text{TMP}$ for given TMP. This should be contrasted with the theoretical UF result for hard spheres and for unchanged operating conditions (see insets of Figs. \ref{fig6}(a) and (b)) where $\phi_w$ is enhanced and $v_w$ lowered by two orders of magnitude. This  marked difference can be attributed to the significantly smaller collective diffusion coefficient for hard spheres, giving rise to a significant enrichment of particles at the membrane wall, with osmotic pressure and viscosity values inside the CP layer that are consequently much larger than those of the silica system. 

\begin{figure}[t!]
\begin{center}
\includegraphics[width=1\linewidth]{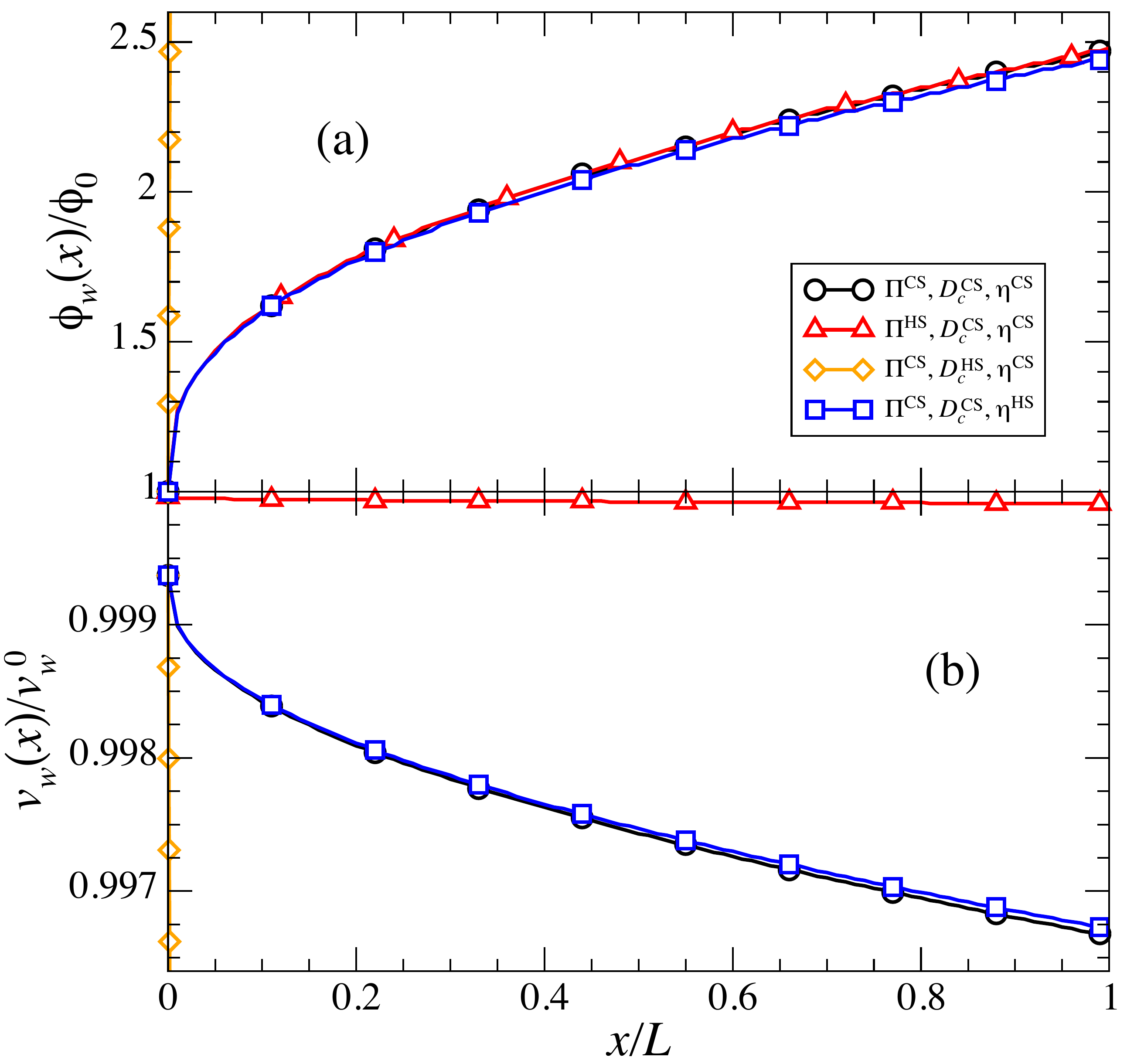}
\caption{(a) Calculated scaled surface concentration profile, $\phi_w(x)/\phi_0$, 
and (b) scaled permeate velocity profile $v_w(x)/v_w^0$. Solid black lines marked by $\circ$ are the results for low-salinity silica suspensions. Red lines marked by $\bigtriangleup$ are results using the osmotic pressure, $\Pi^\text{HS}$, for hard spheres, and the collective diffusion coefficient, $D_c^\text{CS}$, and viscosity, 
$\eta^\text{CS}$, for charged spheres. Orange lines marked by $\Diamond$ are obtained using the collective diffusion coefficient, $D_c^\text{HS}$,  for hard spheres, and the osmotic pressure, $\Pi^\text{CS}$, and viscosity, 
$\eta^\text{CS}$, for charged spheres. Blue lines marked by $\Box$ are obtained using the viscosity of hard spheres, 
and the osmotic pressure and collective diffusion coefficient of charged spheres. System parameters are as in Fig. \ref{fig6}.}
\label{fig7}
\end{center}
\end{figure}

It is instructive to quantify the changes in the CP layer and permeate flux induced by individually replacing $\Pi(\phi)$, $D_c(\phi)$, and $\eta$ of the charged silica particles (CS) by those of neutral hard spheres (HS). This quantification is made in Figs. \ref{fig7}(a) and (b). The black curves marked by $\circ$ are the results for the $\phi_w(x)$ and $v_w(x)$ profiles of the silica system shown in Fig. \ref{fig6}. If in the UF calculation, 
the Carnahan-Starling osmotic pressure for hard spheres is used instead 
of the charged-particles pressure, the red curves marked by $\bigtriangleup$ are obtained. While the CP profile at the membrane remains nearly the same,  
the permeate velocity is now larger and practically equal to $v_w^0$, owing to $\Pi^\mathrm{HS}(\phi) \ll \Pi^\mathrm{CS}(\phi)$ (see Fig. \ref{fig2} and Eq. \ref{eq:Darcy}). If the osmotic pressure and effective viscosity remain those of the silica system but the collective diffusion coefficient of hard spheres is used instead, the orange curves marked by $\Diamond$ are obtained for $\phi_w(x)$  and $v_w(x)$. The CP profile (permeate velocity) is now much larger (smaller)  than that of the silica system, and of values comparatively close to those for a hard-sphere system. This behavior is explained by noting that $D_c^\mathrm{HS}(\phi)\ll D_c^\mathrm{CS}(\phi)$ (see Fig. \ref{fig4}) which gives rise to a strongly reduced  transverse diffusion flux of particles away from the membrane surface and hence to a more pronounced  CP layer. Finally, if the  osmotic pressure and collective diffusion coefficient of the silica spheres system is used in combination with the viscosity of hard spheres (blue curve marked by $\Box$), $\phi_w(x)$ and $v_w(x)$ remain practically equal to the profiles for the original silica system. Thus, the UF performance for the investigated low-salinity suspensions is rather insensitive to changes in the CP layer viscosity.   
\begin{figure}[t!]
\begin{center}
\includegraphics[width=1\linewidth]{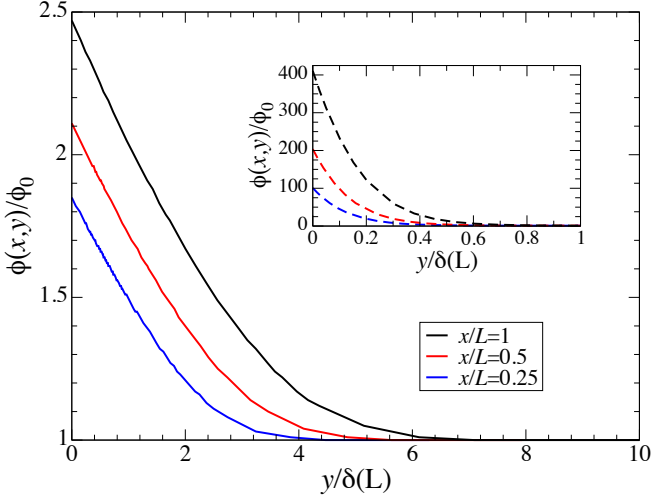}
\caption{Calculated transverse CP layer profile, $\phi(x,y)$, at three different axial positions, $x$, as indicated (differently colored curves), for the low-salinity silica suspension (solid curves). The transverse distance, $y$, from the membrane surface is scaled by the CP layer thickness, 
$\delta(L)=\left(3 D_c(\phi_0)L/\dot{\gamma} \right)^{1/3}$, 
at the fiber outlet. 
Inset: as in main figure, 
but for neutral hard spheres (dashed curves). 
System parameters as in Fig. \ref{fig6}.}
\label{fig8}
\end{center}
\end{figure}

In Fig. \ref{fig8}, the calculated transverse concentration profile, $\phi(x,y)$, inside the CP layer is shown as  function of distance $y$ from the membrane surface, for three different axial positions $x/L$. 
For a comparison, the transverse CP profiles of the hard-sphere system (dashed curves) are displayed in the inset. The transverse profiles decay strictly monotonically from the membrane surface value, $\phi_w(x)$, at $y=0$ down to the feed value, $\phi_0$ that is reached for $y\gg \delta(x)$. Owing to the gradually strengthening CP layer along the membrane surface, $\phi(x,y)$ at a given $y$ is larger for a larger axial distance $x$ from the inlet. Consistent with the axial surface profiles shown in Fig. \ref{fig6}, the transverse profiles of hard spheres depicted in the inset (dashed curves) are two orders of magnitude larger than those of the silica system. 

\subsection{Comparison with UF measurements}

The UF model results are compared here with the outcome of our silica suspension 
cross-flow UF measurements described in Sec. \ref{sec:exp}. 
\begin{figure}[t!]
\begin{center}
\includegraphics[width=1\linewidth]{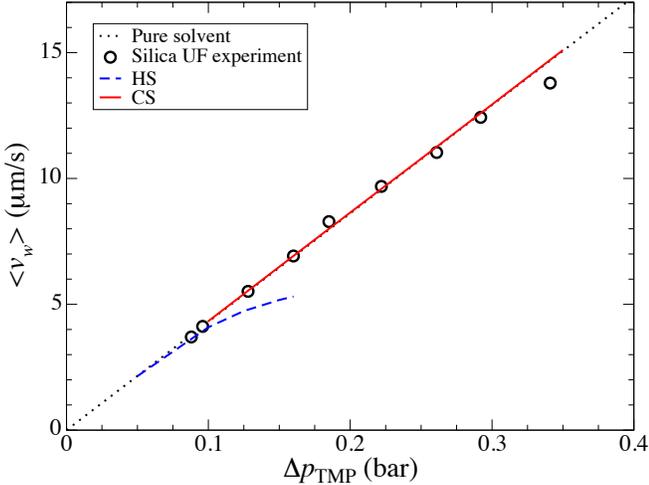}
\caption{Open circles: Our experimental UF results for the fiber-length-averaged permeate velocity, 
$\langle v_w\rangle$, as function of the transmembrane pressure $\Delta p_\mathrm{TMP}$. 
The dotted curve is the result for pure water as feed where $v_w^0=L_p^0\;\!\Delta p_\mathrm{TMP}$. 
System parameters are as in Fig. \ref{fig6}.}
\label{fig9}
\end{center}
\end{figure}

In Fig. \ref{fig9}, the fiber-length-averaged permeate velocity, $\langle v_w \rangle$, is plotted as function of $\Delta p_\mathrm{TMP}$, with the TMP determined using Eq. (\ref{tmp}) from the measured pressure values at the feed, retentate and permeate positions. The dotted black curve is the pure water filtration result, $\langle v_w\rangle=L_p^0\;\!\Delta p_\mathrm{TMP}$, where $L_p^{0}=155$ LMH/bar (cf. Sec. \ref{sec:exp}). The open circles are the experimental findings obtained using Eq. (\ref{vwav_exp}). From our theoretical predictions for $v_w(x)$, $\langle v_w\rangle$ is calculated using
\begin{equation}\label{vwaverage}
\langle v_w\rangle = \frac{1}{L}\int_0^L v_w(x) dx \;\!.
\end{equation}

The dashed blue curve in Fig. \ref{fig9} is the theoretical result for no-slip hard spheres. The influence of the, 
for hard spheres, well developed CP layer becomes visible for $\Delta p_\text{TMP} >0.1$ bar where $\langle v_w\rangle$ is reduced below the straight line characteristic of a pure water feed. The hard-sphere curve of $\langle v_w\rangle$ is truncated at $\Delta p_\text{TMP}\approx 0.17$ where the random closed packing membrane surface concentration $\phi_w=0.64$ is reached, and an amorphous cake layer of jammed particles is formed. In the present work, cake layer formation and other membrane fouling mechanisms are not considered in the theoretical model. 
For charge-stabilized particles  
and charged membranes, this requires an elaborate modeling outside 
the scope of the present work. The solid red curve is the theoretical prediction for 
charged silica particles, obtained by the theoretical methods described in Sec. \ref{sec:model}.

Full agreement is observed in Fig. \ref{fig9} between the experimental and 
theoretical $\langle v_w\rangle$ predictions for the silica system, except for the experimental 
data point for the largest experimental TMP value. The experimental and theoretical data points are close to the pure solvent curve, showing that the osmotic pressure influence is insignificant for the UF of low-salinity suspensions. 
While this finding is surprising on first sight, it was shown in Subsec. \ref{sec:restheo} 
that the strong transverse diffusion flux away from the membrane surface renders the CP layer to be only weakly developed. The reason why the experimental data point for the largest TMP value is 
below the theoretical straight line is definitely not cake formation by crystallization or vitrification. This fouling mechanism is ruled out since the particle concentration values at the membrane surface (cf. Fig. \ref{fig6}) are way too small for the structure factor peak height, $S(q_m;\phi_w)$, to reach the Hansen-Verlet criterion value of $3.1$ 
for low-salinity charge-stabilized systems where the suspension begins to crystallize (see again Fig. \ref{fig3} and Ref. \cite{Gapinski:2014fn}). The deviation of the experimental data point at the largest considered TMP can be attributed  instead to a preferential adsorption of silica particles at the membrane surface. This fouling mechanism is mechanically reversible here, 
since the same value for $L_p^0$ is measured after the silica filtration experiment and a backwashing cycle. The largest-TMP data point can be accounted for in Eq. (\ref{eq:Darcy}) by adding a fouling layer resistance, $R_\text{foul}$, to the membrane resistance according to
\begin{equation}
L_p=\frac{1}{\eta_0(R_\mathrm{mem}+R_\mathrm{foul})}\,.
\end{equation}
Note again that different from the silica system, the osmotic pressure $\Pi^\text{HS}(\phi_w)$ of the reference hard-sphere system contributes significantly in Darcy's law in Eq. (\ref{eq:Darcy}), 
lowering consequently the permeate velocity well below its pure solvent value.

\section{Conclusions}\label{sec:conclusions}

We have undertaken a comprehensive theoretical-experimental study of cross-flow UF of suspensions of charge-stabilized colloidal spheres, for the theoretically challenging case of low-salinity systems where the colloid effective pair potential is distinctly concentration dependent.  
 
The filtration measurements of the integral permeate flux were made using well-characterized aqueous suspensions of charged silica spheres, and a specially designed filtration device. 
The calculations of the axially resolved CP profile, $\phi(x,y)$, 
and permeate velocity, $v_w(x)$, are based on a boundary layer analysis of the coupled   
diffusion-advection and Stokes equations, Darcy's law incorporating 
the influence of the membrane by means of its hydraulic resistance, and the one-component macroion fluid model (OCM) 
of effective colloid interactions. In the OCM description,  
we account for the strong influence at lower salinity of surface-released counterions on the renormalized particle charge and electric screening length, and most importantly on the osmotic pressure $\Pi$. We showed that if $\Pi_\text{OCM}$ alone is used as an approximation of $\Pi$, as done, e.g., Ref. in \cite{Bhattacharjee:1999cz}, the osmotic pressure is severely underestimated. The static pair functions $g(r)$ and $S(q)$, and $\Pi$ were calculated by employing 
the PB combined with the HNC integral equation scheme, and  
used in our calculation of the collective diffusion coefficient and steady-shear suspension viscosity with HIs included. 

We showed that there is a strong electro-hydrodynamic enhancement 
both of $D_c$ and $\eta$ relative to their values for neutral hard spheres, and we have pointed to the invalidity for lower salinity systems of two generalized Stokes-Einstein relations invoking the proportionality of $D_c$ and $1/\eta$.  The good accuracy of the BM-PA and simplified MCT methods of calculating $D_c$ and $\eta$ for charge-stabilized dispersions 
was assessed already in earlier works by the comparison with Stokesian dynamics simulation and experimental results for  colloidal particle suspensions \cite{Heinen:2011if,Westermeier:2012jk}. 

Electrokinetic effects due to the non-instantaneous dynamic response of microion clouds are stronger for small proteins than for the here considered larger colloids, owing to the smaller protein-microion size ratio (see, e.g. Ref. \cite{Gapinski:2005cy}). Particle-specific 
chemical surface charge regulation does not alter the generic behavior of the UF permeate flow 
for low-salinity feed dispersions since its effect is to only moderately increase 
the already quite large renormalized particle charge \cite{Roa:2011gt}. 
Considering the less sophisticated and less accurate transport coefficient approximations used in earlier UF calculations for charge-stabilized dispersions and protein solutions (see, e.g., Refs. \cite{Bowen:2007ix,Bhattacharjee:1999cz}), 
the present study is a significant advancement.   

Additionally to the osmotic pressure, the transport coefficients $D_c$ and $\eta$ are salient input to the macroscopic UF calculations. Our  calculations predict collective diffusion to be of dominant influence. The CP layer is consequently only weakly developed, and it has practically no effect on the (fiber-length-averaged) permeate flux. 
On first sight this is an unexpected result, given the larger values both of $\Pi$ and $\eta$ in comparison to those for uncharged particle systems. While a larger viscosity adds to the CP layer buildup, a larger $\Pi(\phi_w)$ has two antagonistic effects: On the one hand, it lowers $v_w$ according to Darcy's law. 
On the other hand, it lowers $\phi_w$ by indirectly promoting collective back diffusion by means of the 
accordingly lowered osmotic compressibility. We have assessed 
the individual CP layer and permeate flux contributions triggered by $D_c(\phi)$, $\Pi(\phi)$, and $\eta(\phi)$, respectively,  
finding the viscosity to be the least influential one. We have obtained from our calculations that the concentration at the membrane wall is only moderately enlarged by about $8\%$ at the fiber outlet if a viscosity value is used six times larger than the actual one.  
The theoretical results are quantitatively confirmed by our UF experiments showing a 
linear $\langle v_w \rangle$ versus $\Delta p_\text{TMP}$ dependence that coincides practically 
with that for clean water as the feed. 
This is consistent with an earlier related observation by Cohen and Probstein \cite{CohenProbstein:1986} made in the context of reverse osmosis that there is a threshold permeate flux below which no flux decline caused by CP or cake layer formation occurs (see also Ref. \cite{Bacchin:2006}). 

In face of the calculated small $\phi_w$ values at the membrane surface, 
the experimentally observed sub-linearly increasing permeate flux for the largest TMP 
is not explainable by cake formation due to surface crystallization or vitrification. Instead, it is likely due to preferential adsorption of silica particles on the membrane surface that according to our measurements 
is mechanically reversible. The study of membrane fouling mechanisms which in general are membrane, colloid and microion specific is outside the scope of the present work focused on realistic calculations for $\Pi$, $D_c$ and $\eta$ in low-salinity systems, and the assessment of their influence on the axially resolved CP layer and permeate flux in UF. In future work, we will extend the present study to charge-stabilized dispersions with varying salt content, and particle charge and size, including also globular protein solutions. Furthermore, 
we intend to refine our macroscopic UF model by the inclusion of fouling models 
where the bulk phase behavior, particle adsorption on 
the membrane, and membrane pores clogging is accounted for. This will enable us to establish contact, in particular, with a recent  experimental study of fouling processes in a microfluidic filtration setup \cite{Linkhorst:2016}. 

Our method of calculating CP and permeate flux profiles of charge-stabilized systems can be used for optimizing the UF process based on quantitative efficiency criteria for filtration output and energy consumption. 
Suitable criteria have been introduced in Ref. \cite{Roa:2015dj}, and discussed in the context of the cross-flow UF of non-ionic microgel suspensions. Note that the silica suspensions analyzed in this work have salient features in common with more complex ionic microgel supensions. Below the overlap concentration, the direct interactions of ionic microgels are likewise describable in terms of an Yukawa-type effective pair potential such as in Eq. (\ref{eq:OCM potential}), 
however with effective charge and screening parameters depending on the microion penetrability of the microgel polymer backbone \cite{Denton:2014um,Colla:2015hk}. In a realistic modeling of ionic microgels filtration, 
one needs further to account for the microgel shape changes under applied strong pressure gradients and shear flow, 
and for the concentration, salinity, pH, and temperature dependence of the microgel size. 
Experimental-theoretical work by the present authors on microgel filtration is in progress.


\begin{acknowledgments}
We thank J. Buitenhuis and M. Brito (Forschungszentrum J\"ulich), R. Casta\~neda-Priego (University of Guanajuato, Mexico), M. Heinen (CalTech, California and University of Guanajuato, Mexico) and N. Boon (Utrecht University, The Netherlands) for helpful discussions. Financial support by the Deutsche Forschungsgemeinschaft (SFB-985, Project B6) is gratefully acknowledged. 
J.R. acknowledges support by the International Helmholtz Research School on Biophysics and Soft Matter (IHRS BioSoft).
\end{acknowledgments}

\bibliographystyle{apsrev4-1}
\bibliography{ufcharge_refs.bib}

\end{document}